\documentclass[prd,twocolumn,superscriptaddress]{revtex4-1}

\usepackage{graphicx, amsfonts, amsmath, amssymb, amscd,  theorem, exscale,
epic, eepic, epsfig}

\usepackage{hyperref} %

\usepackage{makeidx}
\makeindex

\newcounter{Figure}
\theoremstyle{plain}
\theoremheaderfont{\scshape}

\newtheorem{Def}{\bf Definition}

\theorembodyfont{\upshape}

\theoremheaderfont{\scshape\small} \theorembodyfont{\scshape\small}

\newcommand{\real}{ {\mathbb R} }

\newcommand{\slap}{\mbox{$ \triangle \mkern -13mu / \ $}}
\newcommand{\nlap}{\mbox{$ \nabla \mkern -13mu / \ $}}
\newcommand{\dlap}{\mbox{$ div \mkern -13mu / \ $}}
\newcommand{\Dlap}{\mbox{$ D \mkern -13mu / \ $}}

\newcommand{\clap}{\mbox{$ curl \mkern -23mu / \ $}}

\newcommand{\be}{\begin{equation}}
\newcommand{\ee}{\end{equation}}
\newcommand{\bea}{\begin{eqnarray}}
\newcommand{\eea}{\end{eqnarray}}
\newcommand{\beas}{\begin{eqnarray*}}
\newcommand{\eeas}{\end{eqnarray*}}

\newcommand{\Lie}{ {\mathcal L} }

\newcommand{\Lu}{\underline{L}}

\begin{document}

\title{New Effects in Gravitational Waves and Memory}

\author{Lydia Bieri}
\email{lbieri@umich.edu}
\affiliation{Dept. of Mathematics, University of Michigan, Ann Arbor, MI 48109-1120, USA}


\date{\today}




\begin{abstract} 
We find new effects for gravitational waves and 
memory in asymptotically-flat spacetimes of slow decay. 
In particular, we find growing magnetic memory for these general systems. 
These effects do not arise in spacetimes resulting from data with fast decay towards infinity, including data that is stationary outside a compact set. 
The new results are derived for the Einstein vacuum as well as for the Einstein-fluid equations describing neutrino radiation, where 
the neutrino distribution falls off slowly towards infinity. 
Moreover, they hold for other matter and energy fields coupled to the Einstein equations as long as the data obey corresponding decay laws and other conditions are fulfilled. The magnetic memory occurs naturally in the Einstein vacuum regime of pure gravitation, and in the Einstein-matter systems satisfying the aforementioned conditions. 
As a main new effect, we find that there is diverging magnetic memory sourced by the magnetic part of the curvature. In the most extreme case, the magnetic memory, in addition, features a curl term from the neutrino cloud, growing at the same rate. 
Electric memory is diverging as well, sourced by the electric part of the curvature tensor and the corresponding energy-momentum component. Shear (news) adds to the electric memory. 
Moreover, a multitude of lower order terms contribute to both electric and magnetic memory. 
It has been known that for stronger decay of the data, including data that is stationary outside a compact set, gravitational wave memory is finite and of electric parity only. 
The more general scenarios in this article exhibit richer structures displaying the physics of these more general systems. 
We lay open these new structures. 
Further, we identify a range of decay rates for asymptotically-flat spacetimes for which the new effects occur but with different leading order behavior. 
The new effects are expected to be seen in current and future gravitational wave detectors. They have an abundance of applications of which we mention a few in this paper. Applications include exploring gravitational wave sources of the above types, detecting dark matter via gravitational waves and other areas of physics. 
\end{abstract}

\maketitle


\section{Introduction}

We derive new effects for gravitational radiation and 
memory in asymptotically-flat spacetimes of slow decay. 
The most interesting new effect is the growing magnetic memory. 
In particular, we investigate 
the Einstein vacuum and Einstein-null-fluid equations describing neutrino radiation for sources whose distribution decays very slowly towards infinity. In particular, such sources are not stationary outside a compact set. 
The new effects are expected to be seen in current and future gravitational wave detectors.

It has been known that gravitational wave sources that fall off towards infinity like $r^{-1}$ or faster produce memory effects that are finite and of electric type only. See \cite{chrmemory} for data as (\ref{CKid1}), and \cite{lydia4} for data with $O(r^{-1})$ fall-off without any assumptions on the leading order decay. Thus, magnetic memory does not occur for those sources. The latter include sources that are stationary outside a compact set. 

A completely new panorama opens up when considering spacetimes that decay more slowly to Minkowski spacetime towards infinity. In particular, such sources are not stationary outside a compact set. 
We find that magnetic memory arises naturally within the realm of the Einstein vacuum (EV) equations of pure gravitation, as well as for the Einstein-null fluid (ENF) scenarios. This fact is due to the behavior of the more general, slowly decaying spacetimes, investigated in this article. This new magnetic memory as well as its electric counterpart are diverging at the level of the square root of the absolute value of the retarded time. 
The magnetic memory is sourced by the magnetic part of the curvature tensor, the electric memory by the electric part of the curvature plus an integral term involving the shear. The null fluid contributes to the diverging electric memory at highest rate. 
In the ultimate class of solutions, the magnetic memory, in addition, features a curl term from the energy-momentum tensor for neutrinos also diverging at the aforementioned rate, and the integral of the shear term in the electric null memory becomes unbounded. 
Moreover, we find a multitude of lower order terms contributing at diverging and finite levels.

Gravitational waves in General Relativity (GR) are predicted to change the spacetime permanently. This is known as the memory effect. 
It was found in a linearized theory by Ya. Zel'dovich and B. Polnarev \cite{zeldovich}, and derived in the fully nonlinear setting 
by D. Christodoulou \cite{chrmemory}. D. Garfinkle and the present author showed \cite{lbdg3} that these are two different types of memory rather than a ``linear" and a ``nonlinear" version of the same effect. In particular, we showed \cite{lbdg3} that the former by Zel'dovich and Polnarev, called ordinary memory, is 
related to fields that do not go out to null infinity, whereas the latter by Christodoulou, called null memory, is related to fields that do go out to null infinity.

LIGO's detection \cite{ligodetect1} of gravitational waves from a binary black hole merger in 2015, their joint measurements with Virgo of gravitational waves generated in a neutron star binary merger \cite{ligodetect2, ligodetect3}, as well as several events have fueled hope that memory will be observed for the first time in the near future. In  in \cite{Lasky1}, 
P. Lasky, E. Thrane, Y. Levin, J. Blackman and Y. Chen suggest a way to detect gravitational 
wave memory with LIGO. 

Whereas in detectors like LIGO gravitational memory will show as a permanent displacement of test masses, detectors like NANOGrav will recognize a frequency change of pulsars' pulses.

The pioneering works \cite{zeldovich} and \cite{chrmemory} were followed by several contributions \cite{blda1, blda2, braginsky, braginskyg, will, thorne, thorne2, jorg}. In recent years, the field has grown fast, shedding light on various aspects of memory 
\cite{1lpst1, 1lpst2, lbdg1, lbdg2, lbdg3,tolwal1,winicour, Lasky1, strominger,flanagan,favata,BGYmemcosmo1, bgsty1,twcosmo}. In this article, we concentrate on the references that are relevant for the topics under investigation. See the previous works for a more detailed bibliography. 

Garfinkle and the present author derived \cite{lbdg2} the analogues of both memories for the Maxwell equations. These were the first analogues outside GR. The search for analogues of the memory effect in other theories has grown as well. See for instance \cite{strominger}.

Most physical matter- or energy-fields coupled to the Einstein equations contribute to the null memory \cite{lbdg3}. In particular, this holds for the Einstein-Maxwell system \cite{1lpst1}, \cite{1lpst2}, as well as for neutrino radiation \cite{lbdg1} as it occurs in a core-collapse supernova or a binary neutron star merger. 

In \cite{lydia4} the present author showed that for asymptotically-flat (AF) spacetimes approaching Minkowski spacetime at a rate of $r^{-1}$ and faster, memory is of electric parity only. An interesting and ``unusual" example of stress-energy of an expanding shell in linearized gravity was produced by G. Satishchandran and R. Wald \cite{WaldTm1} that gives rise to an ordinary magnetic memory. 

It is interesting to point out the following: Whereas Satishchandran and Wald \cite{WaldTm1} invoked a very special example of stress-energy to produce magnetic memory, we find magnetic memory even in the Einstein vacuum regime of pure gravity (thus without any stress-energy present). We show that this new magnetic memory occurs naturally in slowly decaying AF spacetimes. 

The present author also showed \cite{lydia4} that asymptotically-flat spacetimes falling off towards infinity at a rate of $o(r^{- \frac{1}{2}})$ generate diverging electric memory. 

In \cite{winma2} T. M\"adler and J. Winicour discussed gravitational wave memory from different sources in a linearized setting. 
They demonstrated that homogeneous, source-free gravitational waves coming in from past null infinity carry a magnetic memory, whereas all the other sources considered in their paper give rise to electric memory only, 
extending Winicour's work \cite{winicour}. The results in \cite{winicour}, \cite{winma2} are consistent with the results from the afore-mentioned literature showing that all gravitational wave memory from systems decaying like $r^{-1}$ or faster are of electric parity only. In these settings, the only way that magnetic memory can be produced is by having incoming radiation from past null infinity ``carry it" already. That is, one has to put it in by hand. 

In the present article, we show that the magnetic memory arises naturally for systems that decay more slowly towards infinity. This holds even for the settings without incoming radiation. Thus, we show that these systems produce magnetic memory in the outgoing radiation without any input from past null infinity. In \cite{lydia12} the present author discussed the situations with incoming radiation in detail. In the latter, we gave the full picture of incoming radiation, comprising gravitational waves of various decay properties. We demonstrated, what happens when we turn on and off this incoming radiation. 
Incoming radiation may consist of primordial gravitational waves stemming from the early universe as well as non-cosmological sources. 

In \cite{lydia12} and in the present article we show that magnetic memory occurs naturally, and we find new memory with new structures.

Our new results for the Einstein vacuum equations also hold for neutrino radiation via the Einstein-null-fluid equations, where, in addition, the stress-energy tensor contributes. As a remarkable property of the ultimate class of spacetimes emerges the contribution to magnetic memory from neutrino radiation through a curl term of its stress-energy tensor, that we also derive in the present article.

The results in this paper are based on a rigorous analysis of the dynamics of the gravitational field in \cite{lydia12} and \cite{lydia1, lydia2}.  
In GR, the dynamics of binary black holes, galaxies and other isolated gravitating systems 
are described by solutions of the Einstein equations approaching Minkowski spacetime at infinity, that is asymptotically-flat (AF) solutions. 
These spacetimes have been comprehended in their full nonlinearity in the proofs of 
global nonlinear stability of Minkowski space. 
Hereby, the Einstein equations for small AF initial data produce globally AF spacetimes that are causally geodescially complete (that is without any singularities). 
The pioneering global proof was established by D. Christodoulou and S. Klainerman in \cite{sta}. 
N. Zipser generalized the works \cite{sta} by Christodoulou and Klainerman to the Einstein-Maxwell system \cite{zip}, \cite{zip2}, and the present author in \cite{lydia1}, \cite{lydia2} to the borderline case for the Einstein vacuum equations assuming one less derivative and less fall-off by one power of $r$ than in \cite{sta} obtaining the borderline case in view of decay in power of $r$. 
It is interesting to note that the smallness conditions in these works are needed to obtain an existence result, whereas 
the main behavior along null hypersurfaces towards future null infinity remains largely independent from the smallness. 
This can be checked by taking a double-null foliation near scri. Examine 
the portion of null infinity 
near spacelike infinity such that the past of this portion intersected with the initial spacelike hypersurface lies within the large data region. 
Then find that the asymptotic results still hold for this portion of null infinity. 
Therefore, the rigorous description in \cite{sta, zip, zip2, lydia1, lydia2} of the behavior of the gravitational field at null infinity not only yields insights into the behavior of data that is allowed to be large, such as black holes, but it also serves as a platform from which important new results on gravitational radiation can be derived. There has been a large literature on extensions of the original stability proof under various assumptions. However, as the point of the present article is not that discussion, we concentrate on the results that are relevant for our new effects. The latter build on spacetimes constructed in \cite{lydia1}, \cite{lydia2}. 

In this article, we make use of the outcome of \cite{lydia12}, using asymptotic results from \cite{lydia1, lydia2}. 
We do not need the mathematical methods that led to these results but apply the latter in important physical settings. 
Here, we give a self-contained derivation of the new physical effects. Further, we derive more results on AF spacetimes enjoying a range of decay properties.

\section{Main Structures}
\label{settingresults}

We consider the Einstein vacuum (EV) equations 
\be \label{EV}
R_{\mu \nu} = 0  
\ee 
as well as the Einstein-null-fluid (ENF) equations describing neutrino radiation in GR 
\be \label{ENF}
R_{\mu \nu}  \ = \ 8 \pi \ T_{\mu \nu} 
\ee
in four spacetime dimensions for $\mu, \nu = 0,1,2,3$ and setting the constants $G=c=1$. 
As the $T_{\mu \nu}$ for the null fluid is traceless, 
the full Einstein equations 
\be \label{ET}
R_{\mu \nu} - \frac{1}{2} R g_{\mu \nu}  \ = \ 8 \pi \ T_{\mu \nu} 
\ee
reduce to (\ref{ENF}). 

The twice contracted Bianchi identities give 
\be \label{BianchiG1}
D^{\mu} G_{\mu \nu} \ = \ 0 
\ee
implying 
\be \label{DT*}
D^{\mu} T_{\mu \nu} \ = \ 0 \ . 
\ee

We denote the solution spacetimes by $(M,g)$. 
Here $t$ is a maximal time function, foliating the spacetime into spacelike hypersurfaces $H_t$, and $u$ denotes the optical function yielding a foliation into outgoing null hypersurfaces $C_u$. Let $S_{t,u} = H_t \cap C_u$ be the $2$-surfaces of intersection. 

When using a $t$-foliation of the $4$-dimensional manifold, we refer to the $0$-component as the $t$-component, and denote the spatial parts by Latin letters $a, b, c, \cdots$. Greek letters $\alpha, \beta, \gamma, \cdots$ denote spacetime indices. 
$T$ is the future-directed unit normal to $H_t$, and $N$ the outward unit normal to $S_{t,u}$ in $H_t$. It is $N = a^{-1} \frac{\partial}{\partial u}$ with lapse $a = | \nabla u |^{-1}$.  
Mainly we work with a null frame $e_1, e_2, e_3, e_4$, with $\{ e_A \}, A = 1,2$ denoting a local frame field for $S_{t,u}$, and $e_3 = \underline{L}, e_4 = L$ are a null pair. That is $g(e_4, e_3) = -2$. 
Note that the outgoing null vector field $e_4 = T+N$, the incoming null vector field $e_3 = T-N$. 
Let $\underline{u} := - u + 2r$, where $r = r(t,u)$ is defined by the surface area of $S_{t,u}$ being $4 \pi r^2$ . Let $\tau_- := \sqrt{1 + u^2}$, and let $\tau_+ := \sqrt{1 + \underline{u}^2}$. 

Denote by $D$ or $\nabla$, the covariant differentiation on $M$, by $\overline{\nabla}$ or $\nabla$ the one on the spacelike hypersurface $H$. 
Operators on the surfaces $S_{t,u}$ are written with a slash. 
For a $p$-covariant tensor field $t$ tangent to $S$, $D_4t$ and $D_3t$ are the projections to $S$ of the Lie derivatives 
$\Lie_4 t$, respectively $\Lie_3 t$.

The Weyl tensor $W_{\alpha \beta \gamma \delta}$ is decomposed into its electric and magnetic parts defined via 
\begin{eqnarray}
{E_{ab}} := {W_{atbt}} \label{Wel1}
\\
{H_{ab}} := {\textstyle {1 \over 2}} {{\epsilon ^{ef}}_a}{W_{efbt}} \label{Wma1}
\end{eqnarray}
with $\epsilon _{abc}$ being the spatial volume element, related to the spacetime volume element by
${\epsilon_{abc}} = {\epsilon_{tabc}}$. The distance between two objects in free fall obeys the following equation depending on the electric part of the Weyl tensor. Namely, the spatial separation $\Delta {x^a}$ is given by 
\begin{equation}
{\frac {{d^2}\Delta {x^a}} {d{t^2}}} = - {{E^a}_b}\Delta {x^b}
\label{geodev}
\end{equation}

Further, the spacetime curvature at $S_{t,u}$ decomposes into the following components: the symmetric $2$-covariant tensorfields 
$\alpha$, $\underline{\alpha}$, the $1$-forms $\beta$, $\underline{\beta}$, and the scalar functions $\rho$, $\sigma$. For any vectors $X, Y$ tangent to $S_{t,u}$ at a point and $\epsilon$ the area $2$-form of $S_{t,u}$, it is 
$\alpha (X,Y) = R(X, e_4, Y, e_4)$, $\underline{\alpha} (X,Y) = R(X, e_3, Y, e_3)$, 
$2 \beta (X, Y) =  R(X, e_4, e_3, e_4)$, $2 \underline{\beta} (X, Y) =  R(X, e_3, e_3, e_4)$, 
$4 \rho = R(e_4, e_3, e_4, e_3)$, $2 \sigma \epsilon (X, Y) = R(X, Y, e_3, e_4)$. 

In the following, 
we state all the equations for the stress-energy tensor of a null fluid describing neutrino radiation in GR. They hold also in the vacuum case with the $T_{\mu \nu}$ being zero. 

The shears $\hat{\chi}$, $\underline{\hat{\chi}}$ are the traceless parts of the second fundamental forms 
$\chi (X, Y) = g(\nabla_{X} e_4, Y)$, respectively $\underline{\chi} (X, Y) = g(\nabla_{X} e_3, Y)$. Further $\zeta$ denotes the torsion-$1$-form. 
They obey the following equations 
\bea
\dlap \hat{\chi} & = &   - \hat{\chi} \cdot \zeta + \frac{1}{2} (\nlap tr \chi + \zeta tr \chi) - \beta +  4\pi T_{A L} \nonumber \\ 
 \label{codazziT1} \\ 
\dlap \hat{\underline{\chi}} & = &  \hat{\underline{\chi}} \cdot \zeta + \frac{1}{2} (\nlap tr \underline{\chi} - \zeta tr \underline{\chi}) + \underline{\beta} + 
4 \pi T_{A \Lu}   \nonumber \\  \label{codazziT2}
\eea
The Gauss equation reads for the Gauss curvature $K$ of the surfaces $S_{t,u}$: 
\begin{equation} \label{GaussENF}
K=-\frac{1}{4}tr\chi tr\underline{\chi }+\frac{1}{2}\widehat{\chi }\cdot 
\underline{\widehat{\chi }}-\rho \left( W\right) - 2 \pi T_{L \Lu}. 
\end{equation}%
Using (\ref{GaussENF}) the mass aspect function $\mu$ and the conjugate mass aspect function $\underline{\mu}$ read 
\bea
\mu \ & = \ &  - \dlap \zeta + \frac{1}{2} \hat{\chi} \cdot \underline{\hat{\chi}} - \rho (W) - 2 \pi T_{L \Lu} \\ 
\underline{\mu} \ & = \ &   \dlap \zeta + \frac{1}{2} \hat{\chi} \cdot \underline{\hat{\chi}} - \rho (W) - 2 \pi T_{L \Lu} . 
\eea

The second fundamental form $k_{ij}$ of $H_t$ decomposes into 
\beas
k_{NN} &=& \delta \\
k_{AN} &=& \epsilon_A\\
k_{AB} &=& \eta_{AB}  \ \ . 
\eeas
Define
\[
  \theta_{AB} =  \langle \nabla_{A} N, e_B \rangle \ \  \mbox{and} \ \  \underline{\xi}_A  =  \frac{1}{2} g(D_3 e_3, e_A) 
\]   

The second fundamental form $k$ satisfies the equations 
\bea
tr k & = & 0   \label{k1}  \\
(curl \ k)_{ij} & = & H(W)_{ij} + \frac{1}{2} \epsilon_{ij}^{\ \ l} R_{0l}  \label{curlk1}  \\ 
(div \ k )_i & = & R_{0i}  \label{divk1}  \ . 
\eea

Next, let us introduce a helpful concept and notation that we will use later. 
The {\itshape signature} $s$ is defined to be the difference of the number of contractions 
with $e_4$ minus the number of contractions with $e_3$. 
Using the signature we present: 
\begin{Def}
Let $W$ be an arbitrary Weyl tensor and let $\xi$ be any of its null components. 
Let $\Dlap_3 \xi$ and $\Dlap_4 \xi$ denote the projections to $S_{t,u}$ of $D_3 \xi$ and $D_4 \xi$, respectively. 
Define the following $S_{t,u}$-tangent tensors: 
\beas
\xi_3 \ & = & \ \Dlap_3 \xi \ + \ \frac{3-s}{2} tr \underline{\chi} \xi \\ 
\xi_4 \ & = & \ \Dlap_4 \xi \ + \ \frac{3+s}{2} tr \chi \xi  \ \ . 
\eeas 
\end{Def}

The Bianchi equations for $\Dlap_3  \rho$, respectively $\Dlap_3 \sigma$, are 
\bea
&& \ \Dlap_3  \rho \ + \ \frac{3}{2} tr \underline{\chi} \rho \  =  \  \label{TBianchiturho3}  \\
& & \ 
 - \dlap \underline{\beta} 
 - \frac{1}{2} \hat{\chi} \underline{\alpha} \ + \ 
 ( \epsilon  - \zeta) \underline{\beta} \ + \ 
2 \underline{\xi} \beta 
\nonumber  \\ 
 & & + \frac{1}{4} ( D_3 R_{34} - D_4 R_{33} ) \nonumber \\ 
\nonumber \\ 
& & \ \Dlap_3 \sigma \ + \ \frac{3}{2} tr \underline{\chi} \sigma \  =  \  \label{TBianchitusigma3}  \\ 
& & \  - \clap \ \ \underline{\beta}  - \frac{1}{2} \hat{\chi} ^*\underline{\alpha}  + 
\epsilon ^*\underline{\beta} - 2 \zeta ^*\underline{\beta} 
 - 2 \underline{\xi} ^*\beta  
\nonumber  \\ 
   & & + \frac{1}{4} ( D_{\mu} R_{3 \nu} - D_{\nu} R_{3 \mu} ) \epsilon^{\mu \nu}_{ \ \ \ 34} \nonumber 
\eea

\section{Spacetimes of Slow Decay}

We consider the following classes of initial data producing corresponding classes of spacetimes.

\begin{Def} (B) \label{intAFB} 
We define an asymptotically flat initial data set $(H_0, \bar{g}, k)$ 
to be a 
(B) initial data set, if $\bar{g}$ and $k$ are sufficiently smooth 
and 
there exists a coordinate system $(x^1, x^2, x^3)$ in a neighborhood of infinity such that for  
$r = (\sum_{i=1}^{3} (x^i)^2 )^{\frac{1}{2}} \to \infty$, it is:  
\bea
\bar{g}_{ij} -  \delta_{ij}  \ & = & \ 
o_3 \ (r^{- \frac{1}{2}}) \label{afgeng}  \\
k_{ij} \ & = & \ o_2 \ (r^{- \frac{3}{2}})    \ .    \label{afgenk}  
\eea
If in (\ref{afgeng})-(\ref{afgenk}) little $o$ is replaced by big $O$, namely 
$\bar{g}_{ij} -  \delta_{ij} = 
O_3 \ (r^{- \frac{1}{2}})$, and 
$k_{ij} = O_2 \ (r^{- \frac{3}{2}})$, then we call this new set an (A) initial data set. 
\end{Def} 
In \cite{lydia1}, \cite{lydia2}, initial data of the type (\ref{afgeng})-(\ref{afgenk}) are investigated, 
weighted Sobolev norms of appropriate energies are controlled, 
yielding geodesically complete spacetimes. In this article, we consider large data of the above type.  
We call {\bf (A) spacetimes} the solutions of the EV equations (\ref{EV}) constructed from (A) initial data, and {\bf (B) spacetimes} the ones from (B) initial data.  Denote by {\bf (AT) spacetimes}, respectively {\bf (BT) spacetimes}, the solutions of the Einstein-null-fluid equations (\ref{ENF}) describing neutrino radiation for initial data of type (A), respectively type (B), and the corresponding energy-momentum tensor, which {\itshape falls off at slow rates towards infinity} (specified in section \ref{neutrinos}). In particular, such sources are not stationary outside a compact set.

We describe the neutrinos as a null fluid in the Einstein equations (\ref{ENF}) represented via its energy-momentum tensor given by 
$
T^{\mu \nu}  = \mathcal{N} K^{\mu} K^{\nu}
$ 
with $K$ a null vector and $\mathcal{N} = \mathcal{N}(\theta_1, \theta_2, r,  \tau_-)$ a positive scalar function depending on $r$, $\tau_-$, and the spherical variables $\theta_1, \theta_2$. 
It is shown in \cite{lydia12} that as $r \to \infty$ we have 
$
\mathcal{N} = O( r^{-2} \tau^{- \frac{1}{2}} )
$ 
Contract the contravariant tensor $T^{\mu \nu}$ with the metric to obtain the covariant tensor $T_{\mu \nu}$. Thus it is 
$T^{LL} = \frac{1}{4} T_{\underline{L} \underline{L}}$.

\subsection{Behavior of Most Important Quantities}
\label{behavior1}

The rigorous analysis of \cite{lydia1}, \cite{lydia2} yields the following behavior that also holds for large data. In particular, the (B) and (BT) spacetimes obey the following, for (A) and (AT) spacetimes all the little $o$ are replaced by big $O$. 
\bea
\underline{\alpha} \ & = & \ O \ ( r^{- 1} \ \tau_-^{- \frac{3}{2}})   \label{resua1} \\ 
\underline{\beta} \ & = & \ O \ ( r^{- 2} \ \tau_-^{- \frac{1}{2}})  \label{resub1}   \\ 
\rho , \ \sigma , \ \alpha , \ \beta \ & = & \ o \ (r^{- \frac{5}{2}})   \label{resrsab1}   \\ 
\hat{\chi}  \ & = & \ o \ (r^{- \frac{3}{2}})   \label{reshatchi1}   \\ 
\underline{\hat{\chi}}  \ & = & \ O \ (r^{-1} \tau_-^{- \frac{1}{2}})  \label{resuhatchi1}  \\ 
tr \chi - \overline{tr \chi} \ & = & \ O \ (r^{- 2})   \label{restrchi1}  \\ 
\zeta  \ & = & \ o \ (r^{- \frac{3}{2}})  \label{resz1}   \\ 
\underline{\zeta}  \ & = & \ o \ (r^{- \frac{3}{2}})  \label{resuz1}   \\ 
K  - \frac{1}{r^2}  \ & = & \  o \ (r^{- \frac{5}{2}})   \label{resK1} \\ 
tr \chi \ & = & \  \frac{2}{r}  \ + \ l.o.t. \label{trchi**1} \\ 
tr \underline{\chi} \ & = & \  - \frac{2}{r} \ + \ l.o.t.  \label{trchibar**1} 
\eea
The abbreviation $l.o.t.$ stands for ``lower order terms". 
Moreover, we have 
\beas
\theta \ & = & \ 
O(r^{-1} \tau_-^{- \frac{1}{2}}) \\ 
\hat{\eta}  \ & = & \ 
O(r^{-1} \tau_-^{- \frac{1}{2}}) \\ 
\epsilon \ & = & \ o(r^{- \frac{3}{2}})  \\ 
\delta \ & = & \ o(r^{- \frac{3}{2}})  
\eeas

{\bf Limits at null infinity $\mathcal{I^+}$:}  
The spacetimes investigated in this article exhibit the following phenomenon. 
As a result of the proof in \cite{lydia1, lydia2}, several quantities, which are defined locally on the surface $S_{t,u}$, do not attain corresponding limits on a given null hypersurface $C_u$ as 
$t \to \infty$. However, the difference of their values at corresponding points at $S_{u}$ and $S_{u_0}$ tends to a limit. For instance, consider $\hat{\chi}$. It is defined locally on $S_{t,u}$. Recall (\ref{reshatchi1}). 
Even though $r^2 \hat{\chi}$ does not have a limit as $r \to \infty$ on a given $C_u$, the difference at corresponding points on $S_u$ in 
$C_u$ and on $S_{u_0}$ in $C_{u_0}$, joined by an integral curve of $e_3$, does have a limit. 
That is, the said difference attains the limit 
\be \label{limitsconcept1}
\int_{u_0}^{u} \Dlap_3 \hat{\chi} \ du' 
\ee
The part of $\hat{\chi}$ with slow decay of order $o(r^{- \frac{3}{2}})$ is non-dynamical, thus, it does not evolve with $u$. It does not tend to any limit at null infinity $\mathcal{I^+}$. 
Similarly, the components of the curvature that are not peeling have leading order terms that are non-dynamical (and do not attain corresponding limits at $\mathcal{I^+}$). Taking off these pieces gives us the dynamical parts of these (non-peeling) curvature components.

\subsection{Future Null Infinity}
\label{FNIG}

By the results of \cite{lydia12} 
the normalized curvature components $r\underline{\alpha }\left( W\right) $, $r^{2}\underline{\beta }\left(
W\right) $, the normalized energy-momentum components 
$r^2 T_{33}, r^3 T_{43}, 
r^3 T_{AB}, r^4 T_{44}$ 
as well as the derivatives 
$D_A T_{3B}$ 
have limits on $C_u$ as $t \rightarrow + \infty $: 
For {(BT) as well as (AT) spacetimes} it is 
\beas
\lim_{C_{u},t\rightarrow \infty }r\underline{\alpha }\left( W\right)
& = & A_{W}\left( u,\cdot \right) \\ 
\lim_{C_{u},t\rightarrow \infty }\,r^{2}\underline{\beta }\left( W\right)
& = & \underline{B}_{W}\left( u,\cdot \right) \\ 
\lim_{C_{u},t\rightarrow \infty }r^2 T_{33} 
& = &  \mathcal{T}_{33} \left( u,\cdot \right) \ . 
\eeas
For {(AT) spacetimes} we have 
\beas
\lim_{C_{u},t\rightarrow \infty }r^3 T_{43} 
& = & \mathcal{T}_{43} \left( u,\cdot \right), \\ 
\lim_{C_{u},t\rightarrow \infty } r^3 T_{AB} 
& = &  \mathcal{T}_{AB}
 \left(
u,\cdot \right) , \\ 
\lim_{C_{u},t\rightarrow \infty }r^4 T_{44} 
& = &  \mathcal{T}_{44} \left( u,\cdot \right), \\ 
\lim_{C_{u},t\rightarrow \infty }r^3 \nlap_A T_{3B} 
& = &  (\nlap_A T_{3B})^*  \left( u,\cdot \right) \ , 
\eeas
where the limits are on $S^{2}$ and depend on $u$. These limits satisfy 
\beas
\left| A_{W}\left( u,\cdot \right) \right| & \leq &  C\left( 1+\left| u\right|
\right) ^{-3/2}, \\ 
 \left| \underline{B}_{W}\left( u,\cdot
\right) \right| & \leq &  C\left( 1+\left| u\right| \right)^{-1/2}, \\ 
\mathcal{T}_{33} \left( u,\cdot \right) &  \leq &  C\left( 1+\left| u\right|
\right) ^{- \frac{1}{2}}, \\ 
(\nlap_A T_{3B})^*  \left( u,\cdot \right) 
 & \leq &  C\left( 1+\left| u\right|
\right) ^{- \frac{1}{2}} \ . 
\eeas

Define 
\bea
Chi_3 & : =  & \lim_{C_u, t \to \infty} \big{(}  r^2 \frac{\partial}{\partial u}  \hat{\chi} \label{defChi3*1} \big{)} \\ 
Chi & : =  & \int_u Chi_3 \ du  \label{defChi3*2}
\eea

For all the spacetimes of classes (A), (B), (AT), (BT), 
on each null hypersurface $C_u$, the limit of $r\widehat{{\underline{\chi }}  }$ exists as $%
t\rightarrow \infty $, that is 
\begin{equation*}
- \frac{1}{2} \lim_{C_{u},t\rightarrow \infty }r\widehat{\underline{\chi }} = 
\Xi \left( u,\cdot \right) 
\end{equation*}
where $\Xi $ is a symmetric traceless $2$-covariant tensor on $S^{2}$ 
satisfying 
\begin{equation*}
\left| \Xi \left( u,\cdot \right) \right| _{\overset{\circ }{\gamma }}\leq
C\left( 1+\left| u\right| \right) ^{-1/2}.
\end{equation*}
Moreover, we have 
\begin{eqnarray}
\frac{\partial \Xi }{\partial u} &=&-\frac{1}{4}A_{W}  \label{XiAT1} \\ 
Chi_3 &=& - \Xi  \label{Sigmau*1} \label{XiChi3T1} \\ 
\underline{B} &=& - 2 \dlap \Xi \label{XiBT1} 
\end{eqnarray}
(\ref{XiBT1}) follows from (\ref{codazziT2}). 
To show that (\ref{XiAT1})-(\ref{XiChi3T1}) hold, we recall from the structure equations 
\bea
\nlap_N \underline{\hat{\chi}} & = & \frac{1}{2} \underline{\alpha} + l.o.t. \\ 
\nlap_N \hat{\chi} & = &  \frac{1}{4} tr \chi \underline{\hat{\chi}} + l.o.t. 
\eea
As far away from the source, that is on every $C_u$ as $t \to \infty$, the lapse $a$ tends to $1$, the following holds asymptotically: 
\bea
\frac{\partial}{\partial u}  \underline{\hat{\chi}} & = & \frac{1}{2} \underline{\alpha} + l.o.t. \label{chiA1} \\ 
\frac{\partial}{\partial u} \hat{\chi} & = &  \frac{1}{4} tr \chi \underline{\hat{\chi}} + l.o.t.   \label{chiuA2}
\eea
(\ref{XiAT1}) follows from (\ref{chiA1}) and (\ref{XiChi3T1}) from (\ref{chiuA2}). 
Moreover, in the case of (AT) spacetimes the following is a direct consequence of (\ref{codazziT2}): 
At future null infinity $\mathcal{I}^+$, it is 
\bea 
- 2 \clap \ \ \dlap \Xi \ & = & \  \clap \ \ \underline{B} \ + \ 4 \pi \big{(} \clap \ \ T  \big{)}^*_{34_3} \ \ \ \ \label{curldivXi1} \\ 
- 2 \dlap \dlap \Xi \ & = & \ \dlap \underline{B} \label{divdivXi1}
\eea

\subsection{Small versus Large Data}

Whereas the smallness assumptions on the data enforce specific behaviors of the main quantities \cite{lydia1, lydia2}, the large data spacetimes, in addition, exhibit a wealth of extra terms. However, the leading order terms as well as a finer structure, established for the small data situations, are also present in the large data spacetimes. In particular, the leading order behavior is rigorous \cite{lydia12}. We will see how this comes into play when investigating gravitational radiation and memory. 

As a consequence of the work \cite{lydia1}, \cite{lydia2}, spacetimes emerging from initial data of the type (\ref{afgeng})-(\ref{afgenk}) for large data as well as for small data display the following structures: 
\bea
\hat{\chi} & = & \{ r^{- \frac{3}{2}} \} + \{ r^{-2} \tau_-^{+ \frac{1}{2}} \}  + l.o.t. \label{hchiterms1} \\ 
\underline{\hat{\chi}} & = &  \{ r^{-1} \tau_-^{- \frac{1}{2}} \} + \{ r^{- \frac{3}{2}} \} + l.o.t.  \label{huchiterms1}
\eea
Denote by $\{ \cdot \}$ the terms of the order given in the brackets. 

Similarly, a consequence, see \cite{lydia12}, of the proof in \cite{lydia1}, \cite{lydia2} is that spacetimes emerging from initial data of the type (\ref{afgeng})-(\ref{afgenk}) under the corresponding smallness conditions enjoy the following structures: 
\bea
\rho & = & \{ r^{- \frac{5}{2}} \} + \{ r^{-3}  \tau_-^{+ \frac{1}{2}}  \} + \{ r^{- 3} \}  \label{rhoterms1} \\ 
& & 
+ \{ r^{-3}  \tau_-^{+ \beta}  \} 
+  O(r^{-3} \omega^{-\alpha})  \nonumber 
\eea
and 
\bea
\sigma & = & \{ r^{- \frac{5}{2}} \} + \{ r^{-3}  \tau_-^{+ \frac{1}{2}}  \} + \{ r^{- 3} \} \label{sigmaterms1} \\ 
& & 
+ \{ r^{-3}  \tau_-^{+ \beta}  \} 
+  O(r^{-3} \omega^{-\alpha})  \nonumber 
\eea
with $\omega$ denoting $r$ or $\tau_-$ and $\alpha > 0$, $0 < \beta < \frac{1}{2}$. 
Moreover, it is 
\[
\rho_3 = O(r^{-3} \tau_-^{- \frac{1}{2}}) 
\]
\[
\sigma_3 = O(r^{-3} \tau_-^{- \frac{1}{2}}) 
\]
(\ref{rhoterms1})-(\ref{sigmaterms1}) and the behavior for $\rho_3$ and $\sigma_3$ are consequences of the proof in \cite{lydia1}, \cite{lydia2} and the 
smallness assumption for the $e_3$-derivatives of $\rho$, respectively $\sigma$ 
\[
\int_H r^4 |\rho_3|^2 \leq c \epsilon 
\]
\[
\int_H r^4 |\sigma_3|^2 \leq c \epsilon 
\]
In particular, $\rho_3$, respectively $\sigma_3$, cannot have any terms of the order $r^{- \frac{5}{2}} \tau_-^{- \frac{3}{2}}$. However, the latter are present in large data spacetimes. 

If we do not assume any smallness, but we allow {\bf large data}, then in addition to the structures (\ref{rhoterms1}), (\ref{sigmaterms1}) there is a multitude of {\bf other terms with various decay behaviors}. In particular, 
$\rho$ as well as $\sigma$ feature terms of the order 
$r^{- \frac{5}{2}} u^{- \alpha}$ with $0 < \alpha$. 

Large data spacetimes exhibit richer structures.

\section{Incoming and Outgoing Radiation}
\label{inoutrad} 

The outgoing radiation is dominated by $\underline{\hat{\chi}}$ and the incoming radiation by $\hat{\chi}$. 

The energy in form of incoming gravitational waves is defined at past null infinity. 
In \cite{DCblh2008}, Demetrios Christodoulou replaced this notion by an integral over advanced time of  
\be \label{en}
e := \frac{1}{2} | \hat{\chi} |^2 
\ee
In \cite{DCblh2008} D. Christodoulou derived the formation of a closed trapped surface and eventually a black hole by the focussing of gravitational waves. The incoming radiation has to be large enough in order to form a black hole. This data obeys a specific hierarchy. In particular, the energy $e$ is of the order $e = O(r^{-2} \underline{u}^{-1})$. 

In this article, we assume that there is no incoming radiation. This constitutes the most physical situation. The new effects at future null infinity are a consequence of the dynamics of the Einstein equations for the very general initial data considered. Nevertheless, in \cite{lydia12} the present author investigated also the case of incoming radiation and found the corresponding effects to hold at past null infinity. 

At future null infinity, let us consider the following quantities. 
For the Einstein vacuum equations we find that 
the energy radiated away per unit angle in a given direction is $F/4\pi$ with 
\be \label{F1}
F(\cdot) = \frac{1}{2} \int_{- \infty}^{+ \infty}  \mid \Xi(u, \cdot) \mid^2  du  \ . 
\ee

For the Einstein-null fluid equations describing spacetimes with neutrinos 
we find that the energy radiated away per unit angle in a given direction is $F_T/4\pi$ with 
\be \label{FXiT1}
F_T(\cdot) = \frac{1}{2} \int_{- \infty}^{+ \infty} \left( \mid \Xi(u, \cdot) \mid^2 \ + \ 2 \pi \ \mathcal{T}_{33} (u, \cdot) \right) du  \ . 
\ee 
In the special class of (AT) spacetimes we find the angular momentum radiated away due to matter is 
\be \label{RcurlT1}
\mathcal{R}_T (\cdot) = 4 \pi \int_{- \infty}^{+ \infty}  \big{(} \clap \ \ T \big{)}^*_{34_3} (u, \cdot)  du  \ . 
\ee

At this point, recall from above (\ref{limitsconcept1}), that 
in the scenarios of our interest, namely spacetimes (B), (A), (BT) and (AT), 
the derivative $\hat{\chi}_3$ takes a well-defined and finite limit at $\mathcal{I}^+$, whereas $\hat{\chi}$ does not.

In the Bianchi equation for $\Dlap_3  \rho$ in the Einstein vacuum case 
\be  \label{Bianchiturho3}
\Dlap_3  \rho \ + \ \frac{3}{2} tr \underline{\chi} \rho \  =  \ 
- \dlap \underline{\beta} 
 - \frac{1}{2} \hat{\chi} \underline{\alpha} \ + \ 
 ( \epsilon  - \zeta) \underline{\beta} \ + \ 
2 \underline{\xi} \beta 
\ee
we focus on the higher order terms, use the notation $\rho_3 : = \Dlap_3  \rho \ + \ \frac{3}{2} tr \underline{\chi} \rho$, and write 
\[
\rho_3 \  =  \ 
 - \underbrace{ \dlap \underline{\beta} }_{= O(r^{-3} \tau_-^{- \frac{1}{2}})}
 - \underbrace{ \frac{1}{2} \hat{\chi} \cdot \underline{\alpha} }_{= O(r^{- \frac{5}{2}} \tau_-^{- \frac{3}{2}})} + \ l.o.t. 
\]
A short computation yields 
\[
\rho_3 \  =  \ 
 - \underbrace{ \dlap \underline{\beta} }_{= O(r^{-3} \tau_-^{- \frac{1}{2}})}
- \underbrace{\frac{\partial}{\partial u} (\hat{\chi} \cdot \hat{\underline{\chi}})}_{ = O(r^{- \frac{5}{2}} \tau_-^{- \frac{3}{2}})} + \underbrace{\frac{1}{4} tr \chi  |\hat{\underline{\chi}}|^2 }_{ = O(r^{-3} \tau_-^{-1})} + \ l.o.t. 
\]
Thus 
\be \label{rho3*}
\rho_3  + \frac{\partial}{\partial u} (\hat{\chi} \cdot \hat{\underline{\chi}})  \  =  \ 
 - \dlap \underline{\beta} + \frac{1}{4} tr \chi  |\hat{\underline{\chi}}|^2 \ =  \ O(r^{-3} \tau_-^{- \frac{1}{2}})
\ee
We will multiply (\ref{rho3*}) by $r^3$ and take the limit on $C_u$ as $t \to \infty$. 
First, we see that each of the leading order terms on the right hand side attains its well-defined limit. 
For the leading order terms on the left hand side the same is true only 
{\bf under the smallness assumptions} of \cite{lydia1, lydia2}. {\bf For large data} there exist additional quantities at leading (and generally at high) order on the left hand side, that do not tend to a limit on scri, but they cancel out. 

{\bf Small data:} If we are given the smallness condition, then on the left hand side of (\ref{rho3*}) it is 
$\rho_3 = O(r^{-3} \tau_-^{- \frac{1}{2}})$ 
and thus $\rho_3$ takes its corresponding limit at future null infinity. 
Moreover, we saw that the structures for $\rho$ (\ref{rhoterms1}) emerge under the smallness conditions from the proof in \cite{lydia1, lydia2}.  
From this and the structures of the involved terms for small data, it follows that $\frac{\partial}{\partial u} (\hat{\chi} \cdot \hat{\underline{\chi}})$ attains its future null limit \cite{lydia12}. Moreover, we have $\int_u \frac{\partial}{\partial u} (\hat{\chi} \cdot \hat{\underline{\chi}}) \ du  = O(r^{-3})$, \cite{lydia12}.

{\bf Large data:} For large data, there are many more terms of order $r^{- \frac{5}{2}} \tau_-^{- \frac{3}{2}}$ 
in 
$\rho_3$ as well as in $\frac{\partial}{\partial u} (\hat{\chi} \cdot \hat{\underline{\chi}})$ 
and potentially terms of order 
$r^{- \frac{5}{2}} \tau_-^{-1 - \alpha}$ with $\alpha \geq 0$ in $\rho_3$. 
The properties of the terms on the right hand side of equation (\ref{rho3*}) enforce cancellations of these terms on the left hand side \cite{lydia12}. 
We take the said limit at $\mathcal{I}^+$ of the left hand side of (\ref{rho3*}), that is after multiplication with $r^3$. Then we find that, after cancellations, the leading order term on the left hand side of (\ref{rho3*}) originates from $\rho_3$ and is of order 
$O(r^{-3} \tau_-^{- \frac{1}{2}})$. 
Thus, upon integration with respect to $u$ this leading order term becomes $O(r^{-3} \tau_-^{+ \frac{1}{2}})$. 
From $\frac{\partial}{\partial u} (\hat{\chi} \cdot \hat{\underline{\chi}})$ there is no contribution at that order. Rather, after cancellations, and subsequent integration with respect to $u$ we derive that the contribution from 
$\int_u \frac{\partial}{\partial u} (\hat{\chi} \cdot \hat{\underline{\chi}}) \ du$ is at the order $O(r^{-3})$.

In $\int_u \frac{\partial}{\partial u} (\hat{\chi} \cdot \hat{\underline{\chi}}) \ du$ the terms of order $O(r^{-3})$ in general depend on $u$. The same is also true (and obvious) for $\rho$.

Comparing the above to asymptotically-flat systems of order $O(r^{-1})$, an eminent difference lies in the fact that for the latter 
the product $(\hat{\chi} \cdot \hat{\underline{\chi}})$ decays in $|u|$. More precisely, one has 
$(\hat{\chi} \cdot \hat{\underline{\chi}}) = O(r^{-3} \tau_-^{- \alpha})$ with $\alpha > 0$. \\

In what follows we consider {\bf large data}.

\section{New Effects}

\subsection{Future Null Infinity and Electric Memory}
\label{FNI}

In this section, we derive the new effects for (B) and (A) spacetimes. 

Building on the results above 
we introduce the following notation for the corresponding limit of the left hand side of (\ref{rho3*}): 
\bea
\mathcal{P}_{3} & : =  & \lim_{C_u, t \to \infty} r^3 \big{(} \rho_3  + \frac{\partial}{\partial u} (\hat{\chi} \cdot \hat{\underline{\chi}}) \big{)}  \label{defrho3*limit**1} \\ 
\mathcal{P} & : =  & \int_u \mathcal{P}_3 \ du \label{defP3*2limit**1} 
\eea
Let us point out that $\mathcal{P}$ is defined on $S^2 \times \real$ up to an additive function $C_{\mathcal{P}}$ on $S^2$, thus the latter is independent of $u$. 
Later, when taking the integral 
$\int_{- \infty}^{+ \infty} \mathcal{P}_3 \ du$, the term $C_{\mathcal{P}}$ will cancel. 

Now, we take the limit of $\big{(}$$r^3$ (\ref{rho3*})$\big{)}$ on $C_u$ as $t \to \infty$. Thereby, 
each term on the right hand side takes a well-defined limit. 
We obtain 
\be \label{Lrho3*}
\mathcal{P}_3  \ = \ 
- \dlap \underline{B} + 2 | \Xi |^2 
\ee
Next, we investigate the deeper structures for $\mathcal{P}$ in (\ref{defP3*2limit**1}). 
From our previous investigations it follows that the leading order term in the limit $\mathcal{P}_{3}$ originates from 
$\rho_3$ only, and therefore its integral over $u$ is the leading order term in $\mathcal{P}$. 
Giving more details, we conclude that $\mathcal{P}$ has the following structure for $0 < \beta < \frac{1}{2}$ and $\gamma > 0$, 
\bea 
\mathcal{P} \ & = & \ \underbrace{\{ \tau_-^{+ \frac{1}{2}} \} \ + \ \{  \tau_-^{\beta} \}}_{ = \mathcal{P}_{\rho_1}}  \ + \ 
\underbrace{\{ \mathcal{F} (u, \cdot) \}}_{= \mathcal{P}_{\rho_2 } - \frac{1}{2} D} \ + \ \{  \tau_-^{-  \gamma} \} \nonumber \\ 
& & \ + \ C_{\mathcal{P}} 
  \label{limitstructures}
\eea
where $\mathcal{F} (u, \cdot) \leq C$. As always $\{ \cdot \}$ denotes terms of the order given inside the brackets. 
All the terms of order $O( \tau_-^{\alpha})$ with $0 < \alpha \leq \frac{1}{2}$ originate from the integral of the limits of the $\rho_3$ part. We denote them by 
$\mathcal{P}_{\rho_1}$. The limit $\mathcal{F} (u, \cdot)$ in (\ref{limitstructures}) consists of the corresponding components of the integral originating from the limits of the terms of order $O(r^{-3})$ in (\ref{rho3*}). 
It has pieces sourced by $\rho_3$ as well as by $\frac{\partial}{\partial u} (\hat{\chi} \cdot \hat{\underline{\chi}})$. 
Denote the former by $\mathcal{P}_{\rho_2 }$ and the latter by $D$. 
Thus (\ref{limitstructures}) reads 
\be \label{limitstructures2}
\mathcal{P} \ = \ \mathcal{P}_{\rho_1} \ + \ \mathcal{P}_{\rho_2 } \ - \frac{1}{2}  \ D \ + \ C_{\mathcal{P}} \ + \ l.o.t. 
\ee
A direct consequence from our analysis above is the fact that $\mathcal{P}_{\rho_2 }$ as well as $D$ depend on $u$.

Next, we use the definitions (\ref{defChi3*1}), (\ref{defChi3*2}), recalling 
\beas
Chi_3 & : =  & \lim_{C_u, t \to \infty} \big{(}  r^2 \frac{\partial}{\partial u}  \hat{\chi} \big{)} \\ 
Chi & : =  & \int_u Chi_3 \ du 
\eeas
as well as equation (\ref{XiBT1})
\[
\underline{B} = - 2 \dlap \Xi 
\]
and equation (\ref{XiChi3T1})
\[
Chi_3 = - \Xi 
\]
Using these in (\ref{Lrho3*}) gives 
\be \label{LCrho3*}
\mathcal{P}_3  \ = \ 
- 2 \dlap \dlap Chi_3 + 2 | \Xi |^2 
\ee
Integrate (\ref{LCrho3*}) with respect to $u$ to obtain 
\be \label{supergold****}
(\mathcal{P}^- - \mathcal{P}^+) - \int_{- \infty}^{+ \infty} | \Xi |^2 \ du 
  \ = \ 
\dlap \dlap (Chi^- - Chi^+) 
\ee
Using (\ref{limitstructures}) and (\ref{limitstructures2}) 
we write 
\bea 
& &
 (\mathcal{P}_{\rho_1 }^- - \mathcal{P}_{\rho_1 }^+)
\ + \ (\mathcal{P}_{\rho_2 }^- - \mathcal{P}_{\rho_2 }^+) 
\ - \  \frac{1}{2} (D^- - D^+)  \nonumber \\ 
& & 
- \int_{- \infty}^{+ \infty} | \Xi |^2 \ du   \nonumber \\ 
& & 
\ = \ 
\dlap \dlap (Chi^- - Chi^+) \label{supergold}
\eea
The last term on the left hand side 
$(- \int_{- \infty}^{+ \infty} | \Xi |^2 \ du)$ is finite, in fact it is borderline within the solutions of \cite{lydia1, lydia2}. 
Thus, it is finite for (B) spacetimes. However, it may not be bounded for general (A) spacetimes. 
We observe that $(\mathcal{P}^- - \mathcal{P}^+)$, respectively $\dlap \dlap (Chi^- - Chi^+)$ are infinite for (B) as well as (A) spacetimes. 
In order to investigate this more precisely, 
recall the concept of limits at null infinity $\mathcal{I^+}$ explained in section \ref{behavior1}. 
Next, we pick $u_0$, then fix a point on the sphere $S^2$ at $u_0$ and consider $\mathcal{P}(u_0)$. At the corresponding point for some $u \neq u_0$ we take $\mathcal{P}(u)$. We keep $u_0$ fixed and let $u$ tend to $+ \infty$, respectively to $- \infty$. 
Then the difference $\mathcal{P}(u) - \mathcal{P}(u_0)$ is no longer finite, but it grows with $|u|^{+ \frac{1}{2}}$ for (A) and (B) spacetimes. 
Taking into account the details of (\ref{supergold}), $(\mathcal{P}_{\rho_1 }(u) - \mathcal{P}_{\rho_1 }(u_0))$ is no longer finite, but it grows with $|u|^{+ \frac{1}{2}}$. 
A corresponding argument holds for $Chi(u) - Chi(u_0)$. 
Finally, in (\ref{supergold}), the second and third terms on the left hand side are finite.

{\bf Remark:} For any asymptotically flat system of the order $O(r^{-1})$, $D$ 
decays in $|u|$ for large $|u|$. More precisely, in those spacetimes, $\hat{\chi}$ as well as $\hat{\underline{\chi}}$ take limits at $\mathcal{I}^+$, and 
the product of their limits vanishes as $|u| \to \infty$. Thus, for those systems it is $D^- = D^+ = 0$.

In the spacetimes investigated here, 
it follows from (\ref{supergold}) that there exists a function $\Phi$ such that 
\bea
 \dlap (Chi^- - Chi^+) & = & \nlap \Phi + X  \label{Phi1a} \\ 
\dlap \dlap (Chi^- - Chi^+) & = & \slap \Phi  \nonumber \\ 
& = & (\mathcal{P} - \bar{\mathcal{P}})^- - (\mathcal{P} - \bar{\mathcal{P}})^+ \nonumber \\ 
& & - 2 (F - \bar{F})  \label{Phi2a} 
\eea
where $X = \nlap^{\perp} \Psi$ for a function $\Psi$ whose Laplacian $\slap \Psi = \clap \ \  \dlap (Chi^- - Chi^+)$. 
For the moment, in order to better display the nature of the new structures, 
we set the contribution from $\Psi$ to zero. 
Immediately afterwards, we treat the most general case with $\nlap^{\perp} \Psi \neq 0$ in section \ref{new} to obtain the complete general result and the full set of equations 
(\ref{Psi1})-(\ref{Phi33}). 

{\bf Remark:} In \cite{chrmemory} Christodoulou showed that for spacetimes as investigated by Christodoulou and Klainerman in \cite{sta} the equation for $\clap \ \ $ is trivial, meaning that there is no contribution from the latter. 
AF initial data in \cite{sta}, on which \cite{chrmemory} is based, takes the following form for $r \to \infty$ 
\bea \label{CKid1} 
\bar{g}_{ij}  &   =  & (1 \ + \ \frac{2M}{r}) \ \delta_{ij}  +  o_4 \ (r^{- \frac{3}{2}})  \label{CKid1}  \\ 
k_{ij} \ & = & \  o_3 \ (r^{- \frac{5}{2}})  \label{CKid2}
\eea
The above (no magnetic memory) is true also for any asymptotically-flat system falling off  towards infinity at order $O(r^{-1})$ without any assumptions on the terms decaying like $r^{-1}$ as shown by the present author in \cite{lydia4}. Therefore, the works \cite{chrmemory} and \cite{lydia4} show that there does not exist any magnetic memory for those systems.

However, the magnetic memory occurs naturally 
in (A) and (B) spacetimes, as we will see in equations (\ref{Psi1})-(\ref{Psi2}) in the system (\ref{Psi1})-(\ref{Phi33}).

Next, we may write system (\ref{Phi1a})-(\ref{Phi2a}) on $S^2$ as 
\bea
 \dlap (Chi^- - Chi^+) & = & \nlap \Phi   \label{Phi1} \\ 
\dlap \dlap (Chi^- - Chi^+) & = & \slap \Phi  \nonumber \\ 
& = & (\mathcal{P}_{\rho_1 } - \bar{\mathcal{P}}_{\rho_1 })^- - (\mathcal{P}_{\rho_1 } - \bar{\mathcal{P}}_{\rho_1 })^+ \nonumber \\ 
& & (\mathcal{P}_{\rho_2 } - \bar{\mathcal{P}}_{\rho_2 })^- - (\mathcal{P}_{\rho_2 } - \bar{\mathcal{P}}_{\rho_2 })^+ \nonumber \\ 
& & - 2 (F - \bar{F})  \label{Phi2} \\ 
& & - \frac{1}{2} (D - \bar{D})^- +  \frac{1}{2} (D - \bar{D})^+ \nonumber
\eea 
This is solved by Hodge theory.

Equations (\ref{Phi1})-(\ref{Phi2}), respectively (\ref{Phi1a})-(\ref{Phi2a}), describe the gravitational wave memory of {\bf electric} parity. It consists of the finite null memory generated by the radiated energy $F$ (finite for (B), unbounded for (A) spacetimes) and the infinite ordinary memory generated by 
$(\mathcal{P} - \bar{\mathcal{P}})^- - (\mathcal{P} - \bar{\mathcal{P}})^+$. More precisely, for fixed $u_0$ the difference $(\mathcal{P}_{\rho_1 } (u) - \bar{\mathcal{P}}_{\rho_1 } (u)) - (\mathcal{P}_{\rho_1 } (u_0) - \bar{\mathcal{P}}_{\rho_1 } (u_0))$ grows like $|u|^{+ \frac{1}{2}}$ as 
$u \to + \infty$, respectively $u \to - \infty$. This memory is due to the gravitational waves radiating. 
Further, the contributions from $(\mathcal{P}_{\rho_2 } - \bar{\mathcal{P}}_{\rho_2 })^- - (\mathcal{P}_{\rho_2 } - \bar{\mathcal{P}}_{\rho_2 })^+$ and 
$- \frac{1}{2}(D - \bar{D})^- +  \frac{1}{2} (D - \bar{D})^+$ are finite. 

The difference $(Chi^- - Chi^+)$ is related to the permanent displacement of test masses in a gravitational wave detector like LIGO. 
That is, $(Chi^- - Chi^+)$ multiplied by a factor including the initial distance of the test masses gives the displacement of the test masses by the ordinary and the null memory. 
For more details we refer to \cite{chrmemory} and \cite{lydia4}.

These results complete the structures of divergent memory derived in \cite{lydia4}.

In addition to the diverging memories, these general spacetimes feature further interesting behavior in 
$(\hat{\chi} \cdot \hat{\underline{\chi}})$ and $(\hat{\chi} \wedge \hat{\underline{\chi}})$, in particular the non-vanishing of the corresponding limits for large $|u|$. 
The integral 
$\int_u \frac{\partial}{\partial u} (\hat{\chi} \cdot \hat{\underline{\chi}}) \ du$ as well as 
$\int_u \frac{\partial}{\partial u} (\hat{\chi} \wedge \hat{\underline{\chi}}) \ du$ 
generates finite electric (former), respectively finite magnetic (latter) memory. The latter is shown in the following section. 
Richer structures affect the behavior of gravitational radiation and memory.

\subsection{Future Null Infinity and New Magnetic Memory}
\label{new}

Next, we treat the most general case with $\nlap^{\perp} \Psi \neq 0$.

It is a well-known fact for AF systems, which are decaying towards infinity like $O(r^{-1})$, that magnetic memory does not occur at all \cite{lydia4}, \cite{chrmemory}. 

However, in the present section, we show that magnetic memory arises naturally in these general spacetimes of slow decay, and that it is growing. First, in the present section, we derive magnetic memory in the realm of the Einstein vacuum equations (\ref{EV}) of pure gravity, then in section \ref{newMneutrinos} for the Einstein-null-fluid system describing neutrino radiation. We find that even further new structures contribute for (AT) spacetimes. 

Next, we are going to derive the new effects for (B) and (A) spacetimes. 

Denote 
$\sigma_3 = \Dlap_3 \sigma + \frac{3}{2} tr \underline{\chi} \sigma$. 
We consider the Bianchi equation for $\sigma_3$ in the Einstein vacuum case 
\[
\sigma_3 \ = \ - \clap \ \ \underline{\beta} - \frac{1}{2} \hat{\chi} \cdot \ ^* \underline{\alpha} + 
\epsilon \ ^* \underline{\beta} - 2 \zeta \ ^* \underline{\beta} - 2 \underline{\xi} \ ^* \beta 
\]
Whereas the lower order terms decay to zero when approaching null infinity, the main part of 
the Bianchi equation reads 
\be \label{s313}
\sigma_3 \ = \ - \clap \ \ \underline{\beta} - \frac{1}{2} \hat{\chi} \cdot \ ^* \underline{\alpha} + l.o.t. 
\ee
We compute 
\be \label{s314}
\sigma_3  + \frac{\partial}{\partial u} ( \hat{\chi} \wedge \hat{\underline{\chi}} ) \ = \ - \clap \ \ \underline{\beta} \ = \ 
O(r^{-3} \tau_-^{- \frac{1}{2}}) 
\ee
From our analysis above it follows that for $\sigma$ the same rules apply as for $\rho$. Moreover, in $\hat{\chi} \wedge \hat{\underline{\chi}}$ the orders of each term are at the level of $\hat{\chi} \cdot \hat{\underline{\chi}}$ above. 

We find that for (B) as well as (A) spacetimes, the following hold: 
\begin{itemize} 
\item[(1)] Highest order terms on the left hand side of (\ref{s314}) cancel. The remaining leading order term originates from $\sigma_3$ and is of order $O(r^{-3} \tau_-^{- \frac{1}{2}})$. 
Multiply the left hand side of (\ref{s314}) by $r^3$ and and take the limit on each $C_u$ for $t \to \infty$. Denote this limit by 
$\mathcal{Q}_3$. 
Thus, 
\be \label{defQ3****2}
\mathcal{Q}_3 : = \lim_{C_u, t \to \infty} r^3 \big{(} \sigma_3  + \frac{\partial}{\partial u} ( \hat{\chi} \wedge \hat{\underline{\chi}} ) \big{)}  
\ee
Let 
\be \label{defQ3*2} 
\mathcal{Q}  : =   \int_u \mathcal{Q}_3 \ du 
\ee
It follows that for fixed $u_0$ the difference $\mathcal{Q} (u) - \mathcal{Q} (u_0)$ grows like $|u|^{\frac{1}{2}}$ as $|u| \to \infty$. 
\item[(2)] We find that $\mathcal{Q}$ has the following structure for $0 < \beta < \frac{1}{2}$ and $\gamma > 0$, 
\bea 
\mathcal{Q} \ & = & \ \underbrace{\{ \tau_-^{+ \frac{1}{2}} \} \ + \ \{  \tau_-^{\beta} \}}_{ = \mathcal{Q}_{\sigma_1}}  \ + \ 
\underbrace{\{ \mathcal{F} (u, \cdot) \}}_{= \mathcal{Q}_{\sigma_2 } - \frac{1}{2}  G} \ + \ \{  \tau_-^{-  \gamma} \}  \nonumber \\ 
& &  \ + \ C_{\mathcal{Q}} 
 \label{limitstructuresdawn}
\eea
where $\mathcal{F} (u, \cdot) \leq C$. Note that 
$C_{\mathcal{Q}}$ is an 
additive function on $S^2$ independent of $u$. Later, when taking the integral 
$\int_{- \infty}^{+ \infty} \mathcal{Q}_3 \ du$, the term $C_{\mathcal{Q}}$ will cancel. 
As a direct consequence from the results above the terms in (\ref{limitstructuresdawn}) of order $O( \tau_-^{\alpha})$ with $0 < \alpha \leq \frac{1}{2}$ originate from the integral of the limits of the $\sigma_3$ part. We denote these terms by $\mathcal{Q}_{\sigma_1}$. 
Moreover, $\mathcal{F} (u, \cdot)$ comprises the corresponding components of the integral originating from the limits of the terms of order $O(r^{-3})$ in (\ref{s314}). 
These go back to $\sigma_3$, respectively to $\frac{\partial}{\partial u} (\hat{\chi} \wedge \hat{\underline{\chi}})$. Denote the former by $\mathcal{Q}_{\sigma_2 }$ and the latter by $G$.

Thus we have 
\be \label{limitstructures2dawn}
\mathcal{Q} \ = \ \mathcal{Q}_{\sigma_1} \ + \ \mathcal{Q}_{\sigma_2 } \ - \frac{1}{2}  \ G  \ + \ C_{\mathcal{Q}}  \ + \ l.o.t. 
\ee
We also find that $\mathcal{Q}_{\sigma_2 }$ as well as $G$ depend on $u$. 
Recall that in AF systems of the order $O(r^{-1})$ 
it is always 
$\hat{\chi} \wedge \hat{\underline{\chi}} = O(r^{-3} \tau_-^{- \alpha})$ with $\alpha > 0$. 
More precisely, in those spacetimes, $\hat{\chi}$ as well as $\hat{\underline{\chi}}$ take limits at $\mathcal{I}^+$, and 
the wedge product of their limits vanishes as $|u| \to \infty$, yielding $G^-=G^+=0$. 
\item[(3)] From the previous point we see that $\sigma$ may have a term of the order $r^{-3}$, which is not possible for AF systems with $O(r^{-1})$ decay. In the latter situations it is always $\sigma = O(r^{-3} \tau_-^{- \alpha})$ with $\alpha > 0$. On the other hand, in these systems it is 
$\rho = O(r^{-3})$. 
\end{itemize}
Now, multiply equation (\ref{s314}) by $r^3$ and take the limit on $C_u$ as $t \to \infty$ to obtain 
\be \label{1supergoldsigma}
(\mathcal{Q}^- - \mathcal{Q}^+) 
  \ = \ 
- \frac{1}{2} \  \clap \ \ \int_{- \infty}^{+ \infty} \underline{B} \ du 
\ee
Using 
(\ref{XiBT1}) and (\ref{XiChi3T1})
we deduce 
\be \label{2supergoldsigmausw}
(\mathcal{Q}^- - \mathcal{Q}^+) 
  \ = \ 
\clap \ \  \dlap (Chi^- - Chi^+) 
\ee
Taking into account (\ref{limitstructures2dawn}) we write 
\bea 
& & \clap \ \  \dlap (Chi^- - Chi^+)   \  =  \ \label{2supergoldsigma} \\ 
& & (\mathcal{Q}_{\sigma_1}^- - \mathcal{Q}_{\sigma_1}^+) 
 \ + \ (\mathcal{Q}_{\sigma_2}^- - \mathcal{Q}_{\sigma_2}^+) 
\ - \  \frac{1}{2} (G^- - G^+)   \nonumber 
\eea
Let us compare this to the system (\ref{Phi1}) - (\ref{Phi2}). 
We recall: in this section, we are considering the general case where $\nlap^{\perp} \Psi \neq 0$. 
A new phenomenon enters the stage. 
In particular, there exist functions 
$\Phi$ and $\Psi$ such that 
$\dlap (Chi^- - Chi^+) = \nlap \Phi + \nlap^{\perp} \Psi$. 
Let $Z := \dlap (Chi^- - Chi^+)$. 
Then the following holds: 
\[
\dlap Z \ = \ \slap \Phi \ \ \ \ , \ \ \ \ \ \clap \ \ Z \ = \ \slap \Psi \ \ . 
\]
The new system reads 
\bea
\dlap (Chi^- - Chi^+) 
 = \nlap \Phi + \nlap^{\perp} \Psi  &  &  \label{Psi1} \\ \nonumber \\ 
  \clap \ \  \dlap (Chi^- - Chi^+) \ = \slap \Psi & &  \nonumber \\ 
= (\mathcal{Q} - \bar{\mathcal{Q}})^- - (\mathcal{Q} - \bar{\mathcal{Q}})^+ & &  \label{Psi2}  \\ 
 \nonumber \\ 
 \dlap \dlap (Chi^- - Chi^+) 
 = \slap \Phi & &  \nonumber \\ 
 =  (\mathcal{P} - \bar{\mathcal{P}})^- - (\mathcal{P} - \bar{\mathcal{P}})^+ & &  \nonumber \\ 
 - 2 (F - \bar{F}) & & \label{Phi33} 
\eea
Note that instead of (\ref{Phi1}) there is (\ref{Psi1}). 
And the $\clap \ $ equation has non-trivial right hand side. 
Taking into account all the details, we write 
\bea
  \dlap (Chi^- - Chi^+)   =    \nlap \Phi + \nlap^{\perp} \Psi  & &  \label{77Psi1**}  \\ \nonumber \\ 
 \clap \ \  \dlap (Chi^- - Chi^+)    =    \slap \Psi & & \nonumber \\ 
   =   (\mathcal{Q}_{\sigma_1} - \bar{\mathcal{Q}}_{\sigma_1})^- - (\mathcal{Q}_{\sigma_1} - \bar{\mathcal{Q}}_{\sigma_1})^+  & & \nonumber \\ 
  + (\mathcal{Q}_{\sigma_2} - \bar{\mathcal{Q}}_{\sigma_2})^- - (\mathcal{Q}_{\sigma_2} - \bar{\mathcal{Q}}_{\sigma_2})^+ & &  \nonumber \\ 
  - \frac{1}{2} (G - \bar{G})^- +  \frac{1}{2} (G - \bar{G})^+ & &  \label{77Psi2**} \\  \nonumber \\ 
 \dlap \dlap (Chi^- - Chi^+)  =  \slap \Phi  & &  \nonumber \\ 
 =  (\mathcal{P}_{\rho_1 } - \bar{\mathcal{P}}_{\rho_1 })^- - (\mathcal{P}_{\rho_1 } - \bar{\mathcal{P}}_{\rho_1 })^+  & & \nonumber \\ 
 (\mathcal{P}_{\rho_2 } - \bar{\mathcal{P}}_{\rho_2 })^- - (\mathcal{P}_{\rho_2 } - \bar{\mathcal{P}}_{\rho_2 })^+  & & \nonumber \\ 
 - 2 (F - \bar{F}) & &  \nonumber \\ 
 - \frac{1}{2} (D - \bar{D})^- +  \frac{1}{2} (D - \bar{D})^+ & & \label{77Phi2}
 \eea
Hodge theory provides the solution to this system. 

We conclude, for both (B) as well as (A) spacetimes, that there is the {\bf new effect} of {\bf magnetic memory} growing with $|u|^{\frac{1}{2}}$ sourced by $\mathcal{Q}$ and finite contributions from both $\mathcal{Q}$ and $G$. Moreover, $\mathcal{Q}$ exhibits further diverging terms at lower order. 
The new magnetic effects emerge in addition to the {\bf electric memory} explained above, and that is growing with $|u|^{\frac{1}{2}}$ sourced by $\mathcal{P}$, and finite contributions from $\mathcal{P}$ and $F$ (the latter may be unbounded for (A)). 

We point out the nature of $\clap \ \ \dlap (Chi^- - Chi^+)$, namely that the right hand side of (\ref{Psi2}), respectively (\ref{77Psi2**}), is non-zero. The new magnetic memories enter through the right hand side of this equation. The right hand side being non-zero is a direct consequence from the more general data investigated here. 
Equations (\ref{77Psi1**})-(\ref{77Phi2}), respectively (\ref{Psi1})-(\ref{Phi33}), show that magnetic memory arises naturally in our more general spacetimes. 
The fact that $\clap \ \ \dlap (Chi^- - Chi^+)$ is non-zero means that there is non-trivial angular momentum. The more general data allow for rotation. 

We re-emphasize that this is not the case for AF systems with fall-off $O(r^{-1})$, where magnetic memory does {\bf not} exist \cite{chrmemory}, \cite{lydia12}. (See second remark in section \ref{FNI} above.)

We conclude that 
gravitational radiation in asymptotically-flat spacetimes of types (A) and (B), thus of slow decay towards infinity, generates the above types of growing memory.

\section{Non-Isotropic Distribution of Neutrinos Source the Radiation} 
\label{neutrinos}

Next, we investigate spacetimes of the types (BT) and (AT) describing neutrino distributions falling off slowly towards infinity, that are modeled by a null fluid coupled to the Einstein equations. Then we derive new effects for these spacetimes.

\subsection{Results and Setting}
\label{neutrinosintro}

How do our new findings relate to situations with a non-trivial energy-momentum tensor? 
In this section, we couple the Einstein equations to a source generating these new phenomena. 

Possible sources for this radiation and the new memory effects are neutrinos moving non-isotropically throughout large regions, resulting in a neutrino cloud with very slow decay. In particular, such a source cannot be stationary outside a compact set.  
 
Neutrinos have tiny mass and move at almost the speed of light. 
In \cite{lbatdgbw} with A. Tolish, D. Garfinkle, and R. Wald we showed that massive particles produce ordinary memory whereas null particles generate null memory; and that ordinary memory due to massive particles with large velocities can mimic the null memory in the limit. 
Thus, describing neutrinos through a null fluid coupled to the Einstein equations gives a good model for astrophysical sources. With D. Garfinkle in \cite{lbdg1} we introduced this model and derived the null memory effect from neutrino radiation for sources in spacetimes falling off like (\ref{CKid1}). As a consequence the stress-energy of these sources falls off fast. 
We showed \cite{lbdg1} that all the memories (of electric type) are finite, and that the neutrinos contribute to the null memory by a finite amount. Moreover, these systems do not generate any magnetic memory. 

How about more general systems? 
The subsequent derivations yield magnetic memory for Einstein-null-fluid systems describing neutrino distributions of slow decay as in 
(BT) as well as (AT) spacetimes. At the same time, it comes out that the equations (\ref{ENF}) 
for data decaying like $O(r^{-1})$ without any assumption on the leading order decaying term, do not produce any magnetic memory.

Next, we are going to show how neutrinos moving non-isotropically and whose distribution decays slowly generate magnetic memory. In particular for 
{\bf (BT) as well as (AT) spacetimes}, we show that these neutrinos cause 
\begin{itemize}                                                    
\item[(a)] a {\bf new magnetic memory} effect {\bf growing like $\sqrt{|u|}$}, sourced by the corresponding {\itshape magnetic part of the curvature} growing at the same rate. 
\item[(b)] an {\bf electric memory} effect {\bf growing like $\sqrt{|u|}$} sourced by the corresponding {\itshape electric part of the curvature} 
and {\itshape the $T_{\underline{L} \underline{L}}$ component of the energy-momentum tensor}, 
each growing at the same rate, and a {\itshape finite contribution from the shear}; 
\end{itemize}
{\bf (AT) spacetimes} in addition 
\begin{itemize} 
\item[(c)] produce a {\bf new} magnetic-type memory via an integral of a curl of $T$ term, growing like $\sqrt{|u|}$. Moreover, none of the memories are bounded for these spacetimes. 
\end{itemize} 
These results are in contrast to systems with neutrino sources whose distribution decays faster towards infinity, including sources that are stationary outside a compact set, where all the memory is finite and of electric parity only \cite{lbdg1}.

We describe the neutrinos as a null fluid in the Einstein equations (\ref{ET}), respectively (\ref{ENF}), represented via its energy-momentum tensor given by 
\be \label{emnf1}
T^{\mu \nu}  = \mathcal{N} K^{\mu} K^{\nu}
\ee
where $K$ is a null vector and $\mathcal{N} = \mathcal{N}(\theta_1, \theta_2, r,  \tau_-)$ a positive scalar function depending on $r$, $\tau_-$, and the spherical variables $\theta_1, \theta_2$. 
The Einstein equations for the general spacetimes, as investigated above, enforce loose decay on the energy-momentum tensor $T^{\mu \nu}$. No symmetry nor other restrictions are imposed. 
The distribution of neutrinos decays very slowly towards infinity, being very non-homogeneous and non-isotropic. 
More precisely, the function $\mathcal{N}$ approaches the following structure towards spacelike infinity, that is for $r \to \infty$: 
\be \label{Auen}
\mathcal{N} = O( r^{-2} \tau^{- \frac{1}{2}} )
\ee
(\ref{Auen}) follows from a straightforward argument using (\ref{DT*}) with $T^{LL}$, and the fact that $div L = tr \chi + l.o.t.$ 


Bursts of neutrinos, such as generated in core-collapse supernovae, initially will run off in all directions, eventually the null direction will dominate. After a very long time, the neutrinos may spread throughout large regions in the universe and formerly dominating flows will taper to slower decays. Thus, we have a ``sea" of neutrinos with slow decay towards infinity and interesting dynamics. In particular, these sources are not stationary outside a compact set. 

Whereas in the case of faster fall-off in \cite{lbdg1} there are strong decay laws satisfied by the stress-energy tensor; 
in the present setting the 
neutrino flow is more general. Nevertheless, the flow obeys a slow convergence to the dominating behavior of the null part $T^{LL}$. 

The present author showed \cite{lydia12} that the components of the energy-momentum tensor have the following decay behavior in {\bf (BT) spacetimes}: 
\beas
T^{LL} \ & = & \ O(r^{-2} \tau_-^{- \frac{1}{2}})  \\ 
T^{AL} \ & = & \  o(r^{- \frac{5}{2}} \tau_-^{- \frac{1}{2}} )  \\ 
T^{L \underline{L}}  \ & = & \  o(r^{-3} \tau_-^{- \frac{1}{2}}) \\ 
T^{AB} \ & = & \ o(r^{-3} \tau_-^{- \frac{1}{2}}) \\ 
T^{A \underline{L}} \ & = & \ o(r^{- \frac{7}{2}} \tau_-^{- \frac{1}{2}}) \\ 
T^{\underline{L} \underline{L}} \ & = & \ o(r^{-4} \tau_-^{- \frac{1}{2}}) 
\eeas
Moreover, it is 
\beas
D_{\underline{L}}T_{A \underline{L}} & = & o(r^{- \frac{5}{2}} \tau_-^{- \frac{1}{2}} )  \\ 
D_B T_{A \underline{L}} & = & o(r^{-3} \tau_-^{- \frac{1}{2}}) \\ 
D_L T_{A \underline{L}} & = & o(r^{- \frac{7}{2}} \tau_-^{- \frac{1}{2}}) 
\eeas
For {\bf (AT) spacetimes} the decay behavior is 
\beas
T^{LL} \ & = & \ O(r^{-2} \tau_-^{- \frac{1}{2}})  \\ 
T^{AL} \ & = & \  O(r^{- \frac{5}{2}} \tau_-^{- \frac{1}{2}} )     \\ 
T^{L \underline{L}}  \ & = & \  O(r^{-3} \tau_-^{- \frac{1}{2}}) \\ 
T^{AB} \ & = & \ O(r^{-3} \tau_-^{- \frac{1}{2}}) \\ 
T^{A \underline{L}} \ & = & \ O(r^{- \frac{7}{2}} \tau_-^{- \frac{1}{2}}) \\ 
T^{\underline{L} \underline{L}} \ & = & \ O(r^{-4}\tau_-^{- \frac{1}{2}}) 
\eeas
Moreover, it is 
\beas
D_{\underline{L}} T_{A \underline{L}} & = & O(r^{- \frac{5}{2}} \tau_-^{- \frac{1}{2}} )  \\ 
D_B T_{A \underline{L}} & = & O(r^{-3} \tau_-^{- \frac{1}{2}}) \\ 
D_L T_{A \underline{L}} & = & O(r^{- \frac{7}{2}} \tau_-^{- \frac{1}{2}}) 
\eeas
The difference between (AT) and (BT) spacetimes will become eminent for the term $D_B T_{A \underline{L}}$ exhibiting the order $O(r^{-3} \tau_-^{- \frac{1}{2}})$ for the former and $o(r^{-3} \tau_-^{- \frac{1}{2}})$ for the latter.

The decay behavior of the components of the energy-momentum tensor follows from the structures of the Einstein equations together with the decay behavior of the relevant quantities in the spacetimes (AT), (BT). They are derived in \cite{lydia12}.

\subsection{New: Growing Electric Memory}

The following holds for (BT) as well as (AT) spacetimes. The only difference is that in the former case the null memory from the shear is finite, but in the latter case unbounded. 

We consider the Bianchi equation (\ref{TBianchiturho3}), concentrating on the highest order terms 
\bea
\Dlap_3  \rho \ + \ \frac{3}{2} tr \underline{\chi} \rho \ & = & \ 
 - \dlap \underline{\beta} 
 - \frac{1}{2} \hat{\chi} \underline{\alpha} 
 - 2 \pi  D_4 T_{33}  \nonumber \\ 
 & &  \  + \ l.o.t. \label{Bianchiturho66} 
\eea
Observe that 
\beas
- \frac{1}{4} D_4 R_{33}  \ & = & \  - 2 \pi  D_4 T_{33} 
\eeas 
Moreover, the leading order term in the last term is given by 
\beas
  - 2 \pi (\Dlap_4 \mathcal{N}) 
 \ & = & \  + 2 \pi tr \chi \mathcal{N}  \ . 
 \eeas 
Thus it is 
\beas
  - 2 \pi  D_4 T_{33} 
  \ & = & \ + 2 \pi tr \chi T_{33} + l.o.t.  
\eeas 
Therefore (\ref{Bianchiturho66}) becomes 
\bea
\rho_3  + \frac{\partial}{\partial u} (\hat{\chi} \cdot \hat{\underline{\chi}})  \  & = &   \ 
 - \dlap \underline{\beta} + \frac{1}{4} tr \chi  |\hat{\underline{\chi}}|^2 \nonumber \\ 
 & &
  + 2 \pi tr \chi T_{33}   + l.o.t.  \nonumber \\ 
  \ & = &   \ O(r^{-3} \tau_-^{- \frac{1}{2}}) 
   \label{Bianchiturho77} 
\eea 
Multiply equation (\ref{Bianchiturho77}) by $r^3$ and take the limit on $C_u$ as $t \to \infty$ to obtain 
\be \label{LrhoT1*}
\mathcal{P}_3  \ = \ 
- \dlap \underline{B} + 2 | \Xi |^2 + 4 \pi \mathcal{T}_{33} 
\ee
The next steps are similar to the corresponding procedure in section \ref{FNI}. Namely, integrate equation (\ref{LrhoT1*}) with respect to $u$, using (\ref{divdivXi1}), which yields 
\bea 
\dlap \dlap (Chi^- - Chi^+)   \  =  \ & & \nonumber  \\ 
(\mathcal{P}^- - \mathcal{P}^+) - \int_{- \infty}^{+ \infty} \big{(} | \Xi |^2  \ + \  2 \pi  \ \mathcal{T}_{33} \big{)} \ du & &   \nonumber  \\  \label{supergoldT1}
\eea
$\mathcal{P}$ features the same structures as in (\ref{limitstructures2}). 
In addition to the behavior already found and described in section \ref{FNI} for the Einstein vacuum equations, we deduce from (\ref{supergoldT1}) and the results in section \ref{FNIG}
that the null memory due to the integral of the null limit $\mathcal{T}_{33}$ of the neutrino distribution grows like $\sqrt{|u|}$. This is different from our results in \cite{lbdg1} in so far that in \cite{lbdg1} the corresponding contribution from neutrino radiation is finite.

\subsection{New: Rotation: Growing Magnetic Memory} 
\label{newMneutrinos}

The next part investigates (AT) spacetimes. The (BT) spacetimes do not exhibit the memory sourced by the curl of $T$, but they do feature all the other memory components. The latter have the same structures. 

We consider the Bianchi equation (\ref{TBianchitusigma3}). Focussing on the highest order terms, we write 
\bea
\Dlap_3 \sigma \ + \ \frac{3}{2} tr \underline{\chi} \sigma \  =  \ & &  \nonumber \\ 
 - \clap \ \ \underline{\beta}  - \frac{1}{2} \hat{\chi} ^*\underline{\alpha}  
 + 4 \pi 
(\clap \ \ T)_{34_3}   \  + \ l.o.t. & & \nonumber \\ 
  \label{Bianchitusigma66}
\eea
That is 
\bea
\sigma_3  + \frac{\partial}{\partial u} ( \hat{\chi} \wedge \hat{\underline{\chi}} ) \ & = & \   \nonumber \\ 
- \clap \ \ \underline{\beta} 
 + 4 \pi 
(\clap \ \ T)_{34_3}   \  + \ l.o.t. 
\ & = & \ 
O(r^{-3} \tau_-^{- \frac{1}{2}})  \nonumber \\ 
  \label{Bianchitusigma77}
\eea
Denote the null limit of $r^3 (\clap \ \ T)_{34_3}$ on $C_u$ as $t \to \infty$ by 
\[ 
(\clap \ \ T)^*_{34_3} = ( (\nlap_A T_{3B})^* - (\nlap_B T_{3A})^* ) \ . 
\]
First, we multiply equation (\ref{Bianchitusigma77}) by $r^3$ and take the limit on $C_u$ as $t \to \infty$ to obtain 
\[
\mathcal{Q}_3  \ = \ 
- \clap  \ \ \underline{B} 
+ 4 \pi \big{(} \clap \ \ T  \big{)}^*_{34_3} 
\] 
In view of (\ref{curldivXi1}) 
this reads 
\be \label{LsigmaT1*}
\mathcal{Q}_3  \ = \ 
2 \ \clap \ \ \dlap  \Xi \ 
+ \ 8 \pi \big{(} \clap \ \ T  \big{)}^*_{34_3} 
\ee 
Next, we integrate equation (\ref{LsigmaT1*}) with respect to $u$ to obtain 
\bea
\clap \ \  \dlap (Chi^- - Chi^+) \ = \ & & \nonumber \\ 
(\mathcal{Q}^- - \mathcal{Q}^+) 
\ + \ 
4 \pi \int_{- \infty}^{+ \infty} \big{(} \clap \ \ T  \big{)}^*_{34_3}  \ du & &  \nonumber \\ 
\label{2supergoldsigmaT1}
\eea 
$\mathcal{Q}$ features the same structures as in (\ref{limitstructures2dawn}). 
In (\ref{2supergoldsigmaT1}) we find a new contribution to the magnetic null memory due to the integral of the null limit $\big{(} \clap \ \ T  \big{)}^*_{34_3}$ of the general neutrino distribution. This new memory grows like $\sqrt{|u|}$. 
We conclude that the new magnetic memory is due to $(\mathcal{Q}^- - \mathcal{Q}^+)$ and $\int_{- \infty}^{+ \infty} \big{(} \clap \ \ T  \big{)}^*_{34_3}  \ du$. 

Next, recall the quantities (\ref{FXiT1}) and (\ref{RcurlT1}). 

Then the new equations for neutrino sources (\ref{supergoldT1}) and (\ref{2supergoldsigmaT1}) give rise to the following system 
\bea
 \dlap (Chi^- - Chi^+) & = & \nlap \Phi + \nlap^{\perp} \Psi  \label{TPsi1} \\ 
\clap \ \  \dlap (Chi^- - Chi^+) & = & \slap \Psi  \nonumber \\ 
& = & (\mathcal{Q} - \bar{\mathcal{Q}})^- - (\mathcal{Q} - \bar{\mathcal{Q}})^+ \nonumber \\ 
& & + (\mathcal{R}_T - \bar{\mathcal{R}}_T)    \label{TPsi2} \\ 
\dlap \dlap (Chi^- - Chi^+) & = & \slap \Phi  \nonumber \\ 
& = & (\mathcal{P} - \bar{\mathcal{P}})^- - (\mathcal{P} - \bar{\mathcal{P}})^+ \nonumber \\ 
& & - 2 (F_T - \bar{F}_T)  \label{TPhi33} 
\eea
with the structures (\ref{limitstructures2}), (\ref{limitstructures2dawn}). 

System (\ref{TPsi1})-(\ref{TPhi33}) is solved by Hodge theory. We find various new memory effects of growing and finite order with the structures derived above. The most striking feature is the occurrence of magnetic memory. For both, the electric as well as the magnetic memory, the leading order terms grow at rate $\sqrt{|u|}$. (AT) spacetimes even produce a new contribution to magnetic memory sourced by the integral of the null limit $\big{(} \clap \ \ T  \big{)}^*_{34_3}$, diverging at the same rate.

{\bf Remark:} {\itshape Behavior along $C_u$ and Limits at $\mathcal{I}^+$}: The fact that the energy-momentum tensor component $T_{A \underline{L}}$ in (\ref{codazziT2}) produces a curl contribution in (\ref{curldivXi1}) is unique to the spacetimes with a metric decaying like $O(r^{- \frac{1}{2}})$ towards infinity, thus for (AT) spacetimes. If we assume just a little more decay such as $o(r^{- \frac{1}{2}})$, as for (BT) spacetimes, then the curl of $T$ decays faster and the limiting equation (\ref{curldivXi1}) reduces to 
\be
- 2 \clap \ \ \dlap \Xi \  =  \  \clap \ \ \underline{B}  \label{curldivXi2}
\ee
Note that the divergence on $S$ of the same component $T_{A \underline{L}}$ is of lower order and therefore there is no $T$-term in (\ref{divdivXi1}). 
In fact, equations (\ref{divdivXi1}) and (\ref{curldivXi2}) have been known to hold for sources of faster decay, including those being stationary outside a compact set. 
We point out that they do hold as well for the more general decay as in (BT). However, for the most general class of spacetimes (AT) with a metric decaying like $O(r^{- \frac{1}{2}})$ towards infinity, the curl contribution of $T$ kicks in.

What we have just derived for neutrino sources whose distribution falls off slowly, is fundamentally different 
from the situation studied in \cite{lbdg1}, where a null fluid with stronger decay is shown to have finite electric memory only. For the latter spacetimes \cite{lbdg1} no magnetic memory is possible, because the relevant components of the energy-momentum tensor decay too fast to produce a limit.

\section{Spacetimes: Range of Fall-Off Rates}
\label{range}

We have established the new effects for spacetimes of types (A), (AT) falling-off as $O(r^{- \frac{1}{2}})$ and (B), (BT) falling-off as $o(r^{- \frac{1}{2}})$. 

Next, we are going to answer the following question: 
At what rate of decay do these new effects show? 

Our results from the previous sections yield a leading order divergence at $\sqrt{|u|}$ of the magnetic as well as electric memories for (A), (B), (AT), (BT) spacetimes. At the same time, the results by D. Christodoulou in \cite{chrmemory}, and by the present author in \cite{lydia4} show that data decaying like $O(r^{-1})$ do not produce any magnetic memory and all the memory effects (being of electric parity only) are finite. 

Now, using our derivations from the previous sections, it follows that spacetimes decaying like $O(r^{- 1 + \alpha})$ for $0 < \alpha \leq \frac{1}{2}$ cause magnetic memory of the above types diverging at $|u|^{+ \alpha}$, except for the portion sourced by the curl of stress-energy. The latter starts occurring only at $\alpha = \frac{1}{2}$, namely at $O(r^{ - \frac{1}{2}})$, thus in (AT) spacetimes. The corresponding electric memories diverge at the same rate. 

We conclude that there exists diverging memory of magnetic and electric type for data with a range of fall-off like $O(r^{- 1 + \alpha})$ for $0 < \alpha$. Therefore, sources with specific fall-off rates, lying within this range, produce magnetic as well as electric memories with characteristic growth rates. This information can be used to gain information on these sources.

\section{Conclusions}
\label{conclusions}

We have derived several new memory effects, and we have found 
new structures in gravitational waves. 

The most fascinating new effect is the growing magnetic memory. 
Such a memory does not exist (not even in finite form) for any system decaying at the order of $r^{-1}$ or faster.  
However, it arises naturally in the more general asymptotically-flat spacetimes for the Einstein vacuum equations of pure gravity. Thus, it emerges as a property of gravitation itself. 
Not only does this effect persist in the corresponding spacetimes describing neutrino radiation via the Einstein-null-fluid equations, but also the stress-energy of the neutrino distribution creates a new contribution originating from the curl of stress-energy. 
The results for the Einstein-null-fluid system continue to hold for other types of matter or energy which obey the corresponding decay laws and other conditions. 

Looking at the Einstein vacuum equations, we found that 
(B) and (A) spacetimes exhibit a growing magnetic memory, and that also 
the electric memory diverges at the same rate as the magnetic counterpart. The electric null memory is due to shear that is finite for (B) and unbounded for (A). Coupling the Einstein equations to a null fluid describing the non-isotropic dynamics of neutrinos, whose distribution decays very slowly towards infinity, provides a source for the new phenomena. For (BT) and (AT) spacetimes we find the corresponding memories in the Einstein-null-fluid system for neutrinos. 
In addition, a new feature occurs in so far as the electric null memory sourced by the $T_{33}$ component of the energy-momentum tensor diverges at the highest rate. This is in contrast to systems of stronger fall-off as in \cite{lbdg1}, where the contribution is finite. A completely different and new phenomenon appears for (AT) spacetimes. Namely, the magnetic memory is powered by a new component sourced by the curl of $T$. The contribution of the latter is diverging at the highest rate. 

On top of the leading effects, we find a wealth of finer structures. The leading order, diverging components and all the remaining diverging components in the limits $\mathcal{P}$ and $\mathcal{Q}$ originate from the corresponding curvature parts. They also generate finite memories. Finite memories are also generated by the components of $\mathcal{P}$ and $\mathcal{Q}$ sourced by $\hat{\chi} \cdot \hat{\underline{\chi}}$, respectively $\hat{\chi} \wedge \hat{\underline{\chi}}$. Again, these properties are unique to these slowly decaying spacetimes. 

In this article, we identified a range of decay rates for asymptotically-flat spacetimes for which the new effects occur but with different, characteristic leading order behavior. These results can be used in gravitational wave detectors to gain more insights into these sources. 

In order to study physical situations of matter distributions as the above, we require the no incoming radiation condition at past null infinity $\mathcal{I}^-$. Then we let the initial data evolve under the coupled Einstein-matter equations. These systems generate outgoing radiation that produce the effects derived above. The diverging memories, and especially the magnetic portion, are unique features of these types of spacetimes of slow fall-off. The question of incoming radiation was addressed in \cite{lydia12}. 

These new gravitational wave memories can be used to detect, identify and gain more information about sources. In principle, these new memory effects should be seen in present and future gravitational wave detectors. 

There is a wide range of applications for the new memory effects. One might want to investigate dark matter halos using the new approach.  The present author and D. Garfinkle have ongoing work into this direction. A more direct use of the present results is the following: 
If dark matter (or parts of it) behaves like neutrinos or similar matter that is non-isotropic and non-stationary outside a compact set but decays very slowly, then the new phenomena 
can be used to detect dark matter via gravitational waves.

For what types of other systems may one expect the new effects to be relevant? We find that 
the results for the (BT) and (AT) spacetimes hold for any Einstein-matter system with an energy-momentum tensor obeying the relevant laws. It is crucial that the latter satisfies specific decay laws and other conditions. The dynamics of the coupled Einstein-matter system dictates what is possible in each single case. Clearly, a null field as investigated above, enjoys all these properties. Now, we can investigate the dynamics of corresponding systems but with massive particles. Hereby, we rely on the results \cite{lbatdgbw} that massive particles create ordinary memory whereas null particles create null memory and that ordinary memory due to massive particles with large velocities can mimic the null memory in the limit. 

The diverging leading parts of the new memories are compelling and potentially ``easier" to detect. At the same time, the many finer structures, in particular the finite contributions from the shear interactions, bear interesting information per se and spur further investigations. 

The dynamics in General Relativity are much richer than one might expect from considering the most obvious sources alone. Sources whose distribution decays slowly towards infinity behave very differently from those with stronger fall-off. The latter include sources that are stationary outside a compact set. Our results from \cite{lydia12} as well as from the present article lay open a panorama of new effects and  structures in the more general spacetimes. Moreover, they open up new alleys to further investigate physical systems from a more general point of view.

\section*{Acknowledgments} 

The author thanks Demetrios Christodoulou for useful remarks on a draft of a related article. 
The author thanks the NSF and the Simons Foundation; 
the author was supported by the NSF Grant No. DMS-1811819 
and 
the Simons Fellowship in Mathematics No. 555809.  \\ \\ \\ \\




\begin{thebibliography}{99} 
\bibitem{ligodetect1}
B. P. Abbott {\it et al.} (LIGO Scientific Collaboration and Virgo Collaboration)
, Phys. Rev. Lett. {\bf 116}, 061102 (2016) 
\bibitem{ligodetect2}
B. P. Abbott {\it et al.} (LIGO Scientific Collaboration and Virgo Collaboration)
, Phys. Rev. Lett. {\bf 116}, 241102 (2016) 
\bibitem{ligodetect3}
B. P. Abbott {\it et al.} (LIGO Scientific Collaboration and Virgo Collaboration)
, Phys. Rev. Lett. {\bf 118}, 221101 (2017) 
\bibitem{lydia1} L. Bieri.  
        \begin{itshape} An Extension of the Stability Theorem of the Minkowski Space
in General Relativity. \end{itshape}
        ETH Zurich, Ph.D. thesis.  \textbf{17178}. 
        Zurich. (2007).  
\bibitem{lydia2} L. Bieri.  
        \begin{itshape} Extensions of the Stability Theorem of the Minkowski Space
in General Relativity. Solutions of the Einstein Vacuum Equations. \end{itshape}
        AMS-IP. Studies in Advanced Mathematics. Cambridge. MA. (2009).      
\bibitem{lydia69} L. Bieri.   
 \begin{itshape} An Extension of the Stability Theorem of the Minkowski Space in General Relativity. \end{itshape} 
 Journal of Differential Geometry.  \textbf{86}.  no.1. (2010). 17-70.     
\bibitem{lydia4} L. Bieri.          
    \begin{itshape} Answering the Parity Question for Gravitational Wave Memory. \end{itshape} 
    Phys. Rev. D 98. 124038. (2018).
\bibitem{lydia12} L. Bieri.       
      \begin{itshape} New Structures in Gravitational Radiation.        
        \end{itshape}
   Submitted (2020). https://arxiv.org/pdf/2010.07418.pdf
\bibitem{1lpst1} L. Bieri, P. Chen, S.-T. Yau. 
  \begin{itshape}Null Asymptotics of Solutions of the Einstein-Maxwell Equations in
General Relativity and 
  Gravitational Radiation.\end{itshape}
  Advances in Theor. and Math.Phys.15.4. (2011). 
\bibitem{1lpst2} L. Bieri, P. Chen, S.-T. Yau. 
  \begin{itshape}  The Electromagnetic Christodoulou Memory Effect and its Application to Neutron Star Binary Mergers.   \end{itshape} 
 Class.Quantum Grav. 29, 21, (2012).  
\bibitem{lbdg1} L. Bieri, D. Garfinkle. 
\begin{itshape} Neutrino Radiation Showing a Christodoulou Memory Effect in General Relativity.   \end{itshape} 
Annales Henri Poincar\'e. 23. 14. 329. (2014). (DOI 10.1007/s00023-014-0329-1). 
\bibitem{lbdg3} L. Bieri, D. Garfinkle. 
  \begin{itshape}   Perturbative and gauge invariant treatment of gravitational wave memory.  \end{itshape} 
Phys. Rev. D. 89. 084039. (2014). 
\bibitem{lbdg2} L. Bieri, D. Garfinkle. 
\begin{itshape}  An electromagnetic analog of gravitational wave memory. \end{itshape} 
Class. Quantum Grav. 30. 19. (2013) 195009.  
\bibitem{bgsty1} L. Bieri, D. Garfinkle, S.-T. Yau. 
  \begin{itshape}   Gravitational wave memory in de Sitter spacetime.  \end{itshape} 
  Phys. Rev. D 94. no.6. (2016) 064040
\bibitem{BGYmemcosmo1} L. Bieri, D. Garfinkle, N. Yunes. 
 \begin{itshape} Gravitational wave memory in $\Lambda$CDM cosmology. \end{itshape} 
Classical and Quantum Gravity. 34. 21. (2017). 215002
\bibitem{lbatdgbw} L. Bieri, A. Tolish, D. Garfinkle, R. Wald.  
{\itshape Examination of a simple example of gravitational wave memory.} 
Phys. Rev. D 90. 044060. (2014).  
\bibitem{blda1} L. Blanchet, T. Damour. 
\begin{itshape} Postnewtonian Generation of Gravitational Waves.  \end{itshape} 
Ann.Inst. H. Poincar\'e. Theor. 50. 377. (1989). 
\bibitem{blda2} L. Blanchet, T. Damour. 
\begin{itshape} Hereditary effects in gravitational radiation. \end{itshape} 
Phys.Rev.D 46. 304. (1992). 
\bibitem{braginskyg}
V.B. Braginsky and L.P. Grishchuk, Sov. Phys. JETP, {\bf 62}, 427 (1985) 
\bibitem{braginsky} V.B. Braginsky, K.S. Thorne. 
Nature (London) {\bf 327}, 123. (1987). 
\bibitem{chrmemory}  D. Christodoulou. 
        \begin{itshape} Nonlinear Nature of Gravitation and Gravitational-Wave
Experiments. \end{itshape}
        Phys.Rev.Letters. \textbf{67}. 
        (1991). no.12. 1486-1489. 
\bibitem{chrIV2000}  D. Christodoulou. 
        \begin{itshape} The Global Initial Value Problem in General Relativity. \end{itshape} 
        Proceedings of the 9th Marcel Grossmann Meeting on General Relativity. Rome. Italy. (2000). 
\bibitem{DCblh2008} D. Christodoulou. 
   \begin{itshape} The formation of black holes in general relativity. \end{itshape} EMS Monographs in Mathematics. European Mathematical Society (EMS), Zurich, (2009). MR2488976 (2009k:83010)
\bibitem{sta} D. Christodoulou, S. Klainerman.
        \begin{itshape} The global nonlinear stability of the Minkowski space.
\end{itshape}
        Princeton Math.Series \textbf{41}. 
        Princeton University Press. Princeton. NJ. (1993).   
\bibitem{dam1} T. Damour. 
 \begin{itshape}  Analytical calculations of gravitational radiation.    \end{itshape}
    Proc. 4th Marcel Grossmann Meeting. Part A. (1986). 365.         
   \bibitem{favata}
M. Favata, Class. Quantum Grav. {\bf 27}, 084036 (2010)        
\bibitem{flanagan}
E. Flanagan and D. Nichols, Phys. Rev. D {\bf 92}, 084057 (2015)
\bibitem{jorg}
J. Frauendiener, Class. Quantum Grav. {\bf 9}, 1639 (1992)

\bibitem{fried1} H. Friedrich. 
        \begin{itshape} On the Existence of $n$-Geodesically Complete or Future Complete Solutions 
        of Einstein's Field Equations with Smooth Asymptotic Structure. \end{itshape}
        Comm.Math.Phys. \textbf{107}. (1986). 587-609.  

                   
\bibitem{Lasky1} 
  P.~D.~Lasky, E.~Thrane, Y.~Levin, J.~Blackman and Y.~Chen,
  Phys.\ Rev.\ Lett.\  {\bf 117}, no. 6, (2016)


\bibitem{winma2} T. M\"adler, J. Winicour. 
  \begin{itshape} The sky pattern of the linearized gravitational memory effect. 
         \end{itshape} 
         Classical and Quantum Gravity. \textbf{33}. 17. (2016). 




\bibitem{WaldTm1} G. Satishchandran, R.M. Wald. 
(2019). 
https://arxiv.org/abs/1901.05942




\bibitem{strominger}
A. Strominger and Zhiboedov, JHEP {\bf 1601}, 086 (2016)

\bibitem{thorne}
K. Thorne, in {\it Gravitational Radiation}, eds. N. Deruelle and T. Piran (North Holland, Amsterdam, 1983)

\bibitem{thorne2}
K.S. Thorne, Phys. Rev. D {\bf 45}, 520 (1992)


\bibitem{tolwal1} A. Tolish,  R. M. Wald. 
Phys. Rev. D {\bf 89}, 064008 (2014). 
\bibitem{twcosmo}
A. Tolish and R. M. Wald, Phys. Rev. D {\bf 94}, 044009 (2016)


\bibitem{winicour}
J. Winicour, Class. Quantum Grav. {\bf 31}, 205003 (2014)

\bibitem{will} A. G. Wiseman, C. M. Will. 
Phys. Rev. D {\bf 44}, R2945 (1991).


\bibitem{zeldovich}
Ya.B. Zel'dovich and A.G. Polnarev, Sov. Astron. {\bf 18}, 17 (1974)
\bibitem{zip} N. Zipser. 
        \begin{itshape} The Global Nonlinear Stability of the Trivial Solution of
the Einstein-Maxwell Equations.  \end{itshape}
        Ph.D. thesis. Harvard Univ. Cambridge MA. (2000).          
\bibitem{zip2} N. Zipser.  
        \begin{itshape} Extensions of the Stability Theorem of the Minkowski Space
in General Relativity. - Solutions of the Einstein-Maxwell Equations.
\end{itshape}
        AMS-IP. Studies in Advanced Mathematics. Cambridge. MA. (2009).          
\end{thebibliography}
\end{document}